\documentclass[12pt]{iopart}
\usepackage{graphicx} % Required for inserting images

\expandafter\let\csname equation*\endcsname\relax
\expandafter\let\csname endequation*\endcsname\relax
\usepackage{amsmath, bm}
\usepackage{amssymb}
\usepackage{adjustbox}
\usepackage{mathtools}
\usepackage{booktabs}
\usepackage{siunitx}
\usepackage{multirow}
\usepackage{amsthm}
\usepackage{tikz}
\usepackage[hyphens]{url}
\usepackage[skip=10pt plus1pt, indent=0pt]{parskip}
\usepackage{natbib}
\usepackage[noabbrev, capitalise, nameinlink]{cleveref}
\crefname{appendix}{}{}

\usetikzlibrary{calc}
\usetikzlibrary{positioning}
\tikzset{every picture/.style={/utils/exec={\sffamily}}}
\tikzset{connect with angle rc/.style={to path={%
let \p1=(\tikztostart),\p2=(\tikztotarget),\n1={\y1-\y2} in 
\ifdim\n1=0pt
 -- (\tikztotarget)
\else
 \ifdim\n1>0pt
 -- ++({((\x2-\x1)*0.4},0) [rounded corners]
 -- ++({(\y2-\y1)*cot(-1*#1)},{\y2-\y1})
 -- (\tikztotarget)
 \else
 -- ++({((\x2-\x1)*0.4},0) [rounded corners]
 -- ++({(\y2-\y1)*cot(#1)},{\y2-\y1})
 -- (\tikztotarget)
 \fi
\fi
}}}

\usepackage{verbatim}

\setcitestyle{authoryear, open={(},close={)}}
\begin{document}
%TC:ignore
\title[Future precipitation over High Mountain Asia]{Refined climatologies of future precipitation over High Mountain Asia using probabilistic ensemble learning}

\author{Kenza Tazi $^{1,\, 2,\, *}$, Sun Woo P. Kim$^{3}$,  Marc Girona-Mata$^{1, \,2}$, Richard E. Turner$^{1}$}

\address{$^1$ Department of Engineering, University of Cambridge, Cambridge, United Kingdom}
\address{$^2$ British Antarctic Survey, Cambridge, United Kingdom}
\address{$^3$ Department of Physics, King’s College London, London, United Kingdom}
\address{$^*$ Author to whom any correspondence should be addressed.}

\ead{kt484@cam.ac.uk}

\begin{abstract}
High Mountain Asia (HMA) holds the highest concentration of frozen water outside the polar regions, serving as a crucial water source for more than 1.9 billion people. Precipitation represents the largest source of uncertainty for future hydrological modelling in this area. 
%Future precipitation predictions remain challenging due to complex orography, lack of in situ hydrological observations, and limitations in climate model resolution and parametrisation for this region. To address the uncertainty posed by these challenges, climate models are often aggregated into multi-model ensembles. While multi-model ensembles are known to improve the predictive accuracy and analysis of future climate projections, consensus regarding how models are aggregated is lacking. 
In this study, we propose a probabilistic machine learning framework to combine monthly precipitation from 13 regional climate models developed under the Coordinated Regional Downscaling Experiment (CORDEX) over HMA via a mixture of experts (MoE). This approach accounts for seasonal and spatial biases within the models, enabling the prediction of more faithful precipitation distributions. The MoE is trained and validated against gridded historical precipitation data, yielding 32\% improvement over an equally-weighted average and 254\% improvement over choosing any single ensemble member. This approach is then used to generate precipitation projections for the near future (2036--2065) and far future (2066--2095) under RCP4.5 and RCP8.5 scenarios. Compared to previous estimates, the MoE projects wetter summers but drier winters over the western Himalayas and Karakoram and wetter winters over the Tibetan Plateau, Hengduan Shan, and South East Tibet.
%Lastly, the MoE approach generalises to other climate variables and model ensembles and can be leveraged for model aggregation in other fields.
\end{abstract}


\submitto{\ERL}
\maketitle

%\quickwordcount{main}
%\quickcharcount{main}
%\detailtexcount{main}
%TC:endignore

\section{Introduction}

%For experiments following historical emissions over 20\textsuperscript{th} century, RCMs are relatively inaccurate compared to other regions \citep{sanjay2017downscaled}. For future scenario experiments, they present differences of over \% in seasonal predictions including overall trends (\cite{sanjay2017downscaled}.

\begin{figure}[ht]
    \centering
    \includegraphics[width=\textwidth]{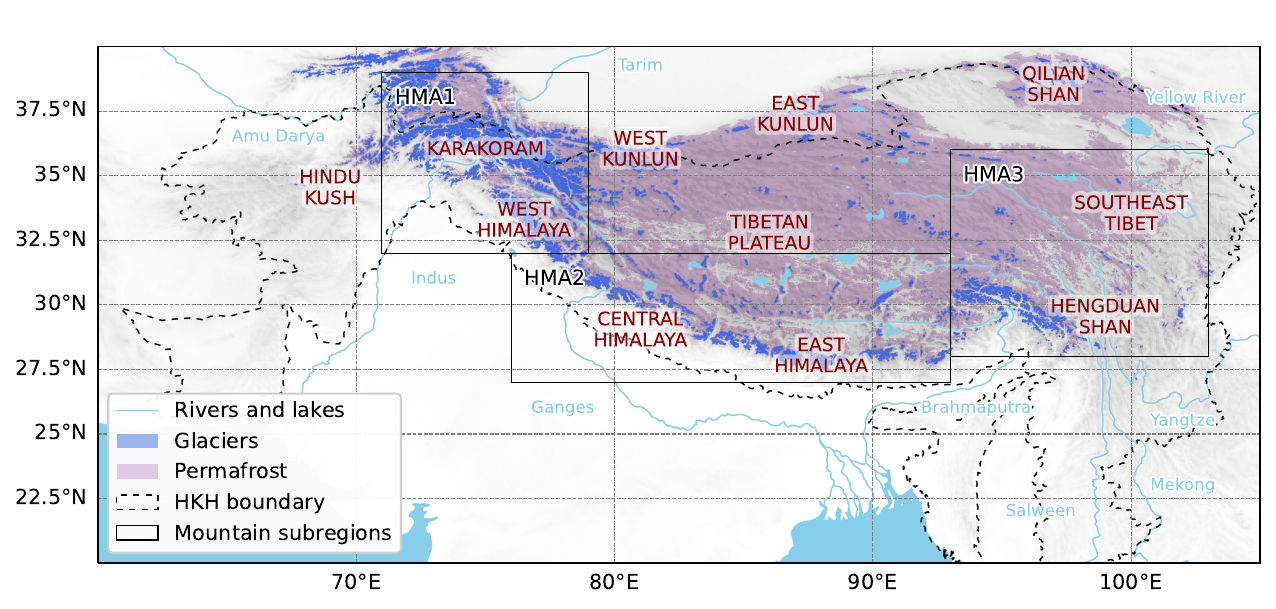}
    \caption[Map of High Mountains Asia]{Map of High Mountain Asia showing major rivers and lakes \citep{lehner2013global}, glaciers \citep{consortium2017randolph}, permafrost \citep{westermann2024esa}, the HKH boundary as defined by \citet{icimod2008outline}, and three mountain subregions used for this study: the West Himalaya and Karakoram (HMA1), the Central and East Himalaya (HMA2), the Hengduan Shan and Southeast Tibet (HMA3). The map also includes the standardised names of the glacierised mountains regions \citep{gtn2017gtn} referenced in this paper.}
    \label{fig:hma_map}
\end{figure}

High Mountain Asia (HMA) stores more frozen water than anywhere else in the world after the Arctic and Antarctic polar caps. Its glaciers and snowfields, shown in \cref{fig:hma_map}, provide fresh water to more than 1.9 billion people through Asia's main rivers \citep{immerzeel2020importance}. Despite the critical importance of these water resources, much remains unknown about their distribution and how they will evolve under climate change. Among all hydrological drivers, precipitation contributes the greatest uncertainty to modelling future water security over HMA \citep{nie2021glacial, orr2022knowledge, wester2023water}. The main tools for understanding future precipitation over HMA are regional climate models (RCMs) \citep{maussion2014precipitation, norris2017spatiotemporal, norris2019deciphering, orr2017sensitivity, norris2020warming, palazzi2015precipitation}. However, complex orography and the lack of in situ hydrological observations, combined with limited RCM resolution and parametrisations over mountains, make precipitation difficult to model in this region \citep{tazi2024downscaling, tazi2024precipitation, orr2017sensitivity,
girona2024probabilistic}. To overcome these challenges and understand the range of possible changes in precipitation, RCMs with different model physics and parametrisations are run, and their outputs combined.

Such collections of climate models are known as multi-model ensembles. In general, multi-model ensembles improve the accuracy of historical predictions and provide a way to explore the uncertainties in model-based climate projections \citep{palmer2005probabilistic,stainforth2007issues, kharin2002climate}. However, the way in which models are combined is controversial \citep{tebaldi2007use}. This includes disagreements about how to measure aggregate accuracy \citep{eyring2019taking}, deal with ensemble outliers \citep{sanderson2012interpretation}, account for the proximity of model design \citep{masson2011climate, knutti2010end}, and address an infinite number of possible ensemble members \citep{tel2020theory}. As a result, the Intergovernmental Panel on Climate Change opted for the unweighted average of the global climate models (GCMs) from the Climate Model Intercomparison Project (CMIP) in their previous reports, using model spread as the uncertainty for future predictions \citep{flato2014evaluation}.\footnote{The sixth report (AR6) discards `hot-models' from CMIP6 that have been shown to be overly sensitive to emission forcing \citep{zelinka2020causes, hausfather2022climate, lee2021future}.}

In this paper, we present a probabilistic machine learning framework with the aim of combining precipitation outputs from 13 RCMs developed under the Regional Downscaling Experiment (CORDEX) for CMIP5. This method leverages the RCMs' heterogeneous seasonal and spatial biases to generate precipitation probability distributions for any location and month. This is achieved through a two-step approach. First, we compute statistical surrogates of the RCMs, i.e. a representation of their simulated precipitation field. For a given climatological period, we compile the empirical precipitation distribution for each RCM, location, and month. We then model the empirical distributions with Gaussian processes (GPs). GPs have been used in previous research for similar problems, such as spatiotemporal aggregation of GCMs using deep kernel learning \citep{harris2013probabilistic} or to find the latent signal of a temperature change in GCM ensembles for a single location \citep{amos2022identifying}.

Second, we aggregate the RCM surrogate predictions at each spatiotemporal point via a mixture of experts (MoE), where distributions of experts (i.e. RCM surrogates) are combined using weights. We parametrise the weight of each expert by their Wasserstein distance with respect to historical data (APHRODITE) and a learnt `statistical temperature'. Our ensemble learning method is therefore similar to Bayesian model averaging \citep{massoud2023bayesian, min2007probabilistic, olson2016simple}, where the similarity of the probability distribution with respect to historical data is used as a measure of faithfulness. However, unlike previous ensembling techniques, our method neither gives each RCM one vote, nor does it completely discard outliers with less accurate distributions; it only down-weighs them. As more experts are added, more precipitation events are considered, and the final distributions are refined.

Our main goal is to compare this method with that of \citet{sanjay2017downscaled}. Their work presented CORDEX CMIP5 runs over HMA. We study the same domain, time periods, and RCMs using both a learnt MoE and an equally weighted (EW) RCM aggregation scheme. The latter is equivalent to the ensembling method used in \citet{sanjay2017downscaled}. The paper is structured as follows. The study area and data are first described in \cref{sec:study_area_data}. The method is then summarised in \cref{sec:method} and validated in \cref{sec:hist_validation}. Predictions for a historical reference period (1976--2005), the near future (2036--2065), and the far future (2066--2095) for RCP4.5 and RCP8.5 are analysed in \cref{sec:refined_clima}. Finally, limitations and further work are discussed in \cref{chap4-sec:discussion}.

\section{Study area and datasets}
\label{sec:study_area_data}

\subsection{Study area}
Following \citet{sanjay2017downscaled}, we study precipitation over HMA between \ang{20}-\ang{40}N \ang{60}-\ang{100}E as shown in \cref{fig:hma_map} and three mountain subregions: the West Himalaya and Karakoram (HMA1, \ang{32}--\ang{39}N \ang{71}--\ang{79}E), the Central and East Himalaya (HMA2, \ang{27}--\ang{32}N \ang{76}--\ang{93}E), and the Hengduan Shan and Southeast Tibet (HMA3, \ang{27}--\ang{32}N \ang{93}--\ang{103}E). In addition to including most glaciers for HMA, these subregions are also representative of key synoptic atmospheric patterns that drive precipitation in this area. Precipitation over HMA1 is strongly influenced by the western disturbances during winter (December to March), while precipitation distributions over HMA2 and HMA3 are mainly driven by the Indian summer monsoon and the East Asian summer monsoon, respectively, from June to September. The study area also overlaps the Hindu Kush Himalayan (HKH) region. As defined by the International Centre for Integrated Mountain Development, the HKH extends from \ang{16} to \ang{39}{N} and \ang{61} to \ang{105}E and encompasses more than \SI{4192000}{\kilo\meter^2} of mountains over Afghanistan, Bangladesh, Bhutan, China, India, Myanmar, Nepal, and Pakistan \citep{bajracharya2011status}.

\subsection{CORDEX-WAS}
CORDEX is a global initiative that aims to provide high-resolution climate data for regional and local applications. CORDEX for the West Asia domain (CORDEX-WAS, \ang{-20}--\ang{50}N \ang{20}--\ang{115}E), also known as the South Asia domain, uses RCMs to dynamically downscale global climate simulations from CMIP5 over most of HMA \citep{sanjay2017downscaled, taylor2012overview}. CORDEX-WAS outputs are chosen over other RCM simulations for their accessibility and rigorous experimental protocol which allow for the straightforward comparison of many models \citep{giorgi2015regional}. CORDEX-WAS is made up of 13 models using three RCMs and ten driving CMIP5 GCMs, as listed in \cref{tab:rcms}. 

%TC:ignore
\begin{table}[ht]

\caption[CORDEX-WAS RCMs with CMIP5 driving models]{CORDEX-WAS RCMs with CMIP5 driving models. Note that GFDL-ESM2M, MPI-ESM-LR, and CNRM-CM5 each drive two RCMs. \dag The SMHI-RCA4 CNRM-CM5 model output only extends to 2085 for RCP8.5. \ddag MPI-REMO2009 MPI-ESM-LR model output only begins from 1961.}
\footnotesize
\begin{center}
\begin{tabular}{ll}
\toprule
\textbf{RCM} & \textbf{Driving CMIP5 model}\\
\midrule
IITM-RegCM4 & CanESM2 \citep{chylek2011observed}\\
\citep{giorgi2012regcm4} & GFDL-ESM2M  \citep{dunne2012gfdl}\\
& CNRM-CM5 \citep{voldoire2013cnrm}\\
& MPI-ESM-MR \citep{jungclaus2013characteristics}\\
& IPSL-CM5A-LR \citep{dufresne2013climate}\\
& CSIRO-Mk3.6 \citep{jeffrey2013australia} \\
\midrule
SMHI-RCA4 & EC-Earth \citep{hazeleger2010ec}\\
\citep{samuelsson2015surface} & MIROC5 \citep{watanabe2010improved}\\
& MPI-ESM-LR  \citep{jungclaus2013characteristics}\\
& IPSL-CM5A-MR \citep{dufresne2013climate}\\
& GFDL-ESM2M \\
& CNRM-CM5\dag \\
\midrule
MPI-REMO2009 & MPI-ESM-LR\ddag \\
\citep{teichmann2013does} & \\
\bottomrule
\end{tabular}\\
\end{center}

\label{tab:rcms}
\end{table}
%TC:endignore

The CORDEX outputs are stored on a \ang{0.44} rotated grid (approximately 50\unit{\kilo\meter} resolution). However, for simplicity, we use interpolated outputs on a \ang{0.5} regular grid.  For precipitation, the Centre for Climate Change Research at Indian Institute of Tropical Meteorology used bilinear interpolation to regrid the data \citep{grid-email}. We analyse CORDEX-WAS monthly precipitation outputs over \ang{20}--\ang{40}N and \ang{60}--\ang{105}E for historical (1950--2005) and projected (2006--2100) scenarios. Historical scenario experiments use estimates of greenhouse gas concentrations, aerosols, and land use (from observed or reconstructed data) to drive the GCMs. Projected scenario experiments are based on representative carbon pathways (RCPs), i.e., theoretical future emissions scenarios. %defined by radiative forcing in watts per square metre by the year 2100. 
In this paper, outputs for RCP4.5 and RCP8.5 are investigated. RCP4.5 represents a stabilisation pathway, where emissions peak around 2040 and then decline, while RCP8.5 represents a high-emissions pathway with minimal policy intervention. Going forward, we simplify simulation names to `GCM RCM', e.g. the IITM-RegCM4 simulation driven by IPSL-CM5A-LR becomes IPSL RegCM4.

\begin{figure}[ht]
    \centering
    \includegraphics[width=\textwidth]{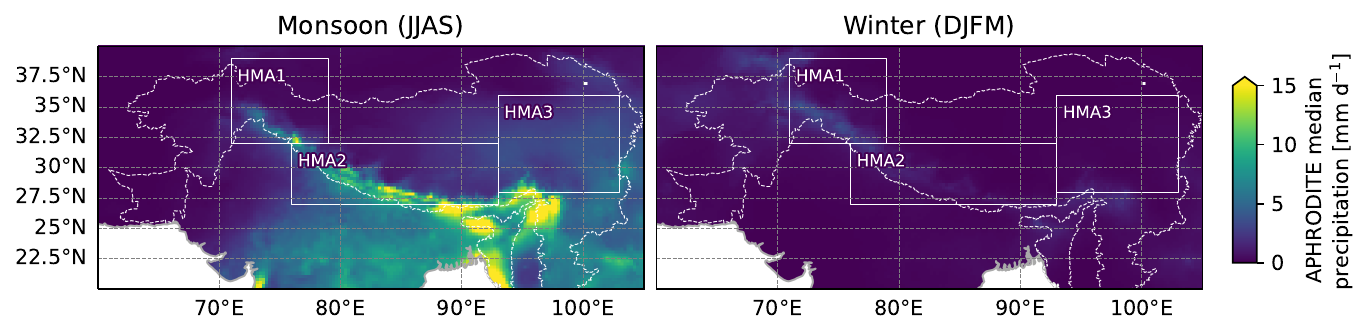}
    \caption[APHRODITE precipitation medians between 1976 and 2005]{APHRODITE precipitation medians between 1976 and 2005 for the summe monsoon (left) and winter (right).}
    \label{fig:aphrodite}
\end{figure}

\subsection{APHRODITE}
\label{sec:aphro}

The Asian Precipitation-Highly Resolved Observational Data Integration Towards Evaluation of Water Resources (APHRODITE) is a gridded precipitation dataset ranging from 1951 to 2015 with a spatial resolution of \ang{0.25} \citep{yatagai2012aphrodite}. APHRODITE has one of the best spatiotemporal coverages over High Mountain Asia and is one of the most studied and accurate precipitation products for the region \citep{dimri2021bias}. The dataset was created through the interpolation of precipitation gauge observations using a custom correlation distance lookup table. We use the monthly precipitation product APHRO\_V1101 specifically developed for monsoon Asia. \cref{fig:aphrodite} plots APHRODITE precipitation between 1976 and 2005.

\section{Method}
\label{sec:method}
This section presents our ensemble learning method, summarised in \cref{fig:bcm_diagram}.

%TC:ignore

\begin{figure}[ht]
    \centering
    \begin{adjustbox}{width=\textwidth} %\includegraphics[scale=0.6]{figures/bcm_diagram.pdf}}
    %\resizebox{330pt}{100pt}{%
    \tikzset{
        io1/.style={rectangle, rounded corners, minimum width=3cm, minimum height=1cm, scale=0.75, text centered, draw=black},
        io2/.style={minimum width=3cm, scale=0.75, minimum height=1cm, text centered},
        arrow/.style={thin, ->, >=stealth}}
        \begin{tikzpicture}[node distance = 3cm]
        
        \node (start) [io1] {CORDEX-WAS};
        
        \node (in1) [io1, right of=start, yshift=1.25cm, xshift=1cm] {RCM 1 ($\mathcal{D}_1$)};
        \node (in0) [io2, below of=in1, yshift=1.75cm] {\rotatebox{90}{...}};
        \node (in11) [io1, right of=in1, yshift=1cm, xshift=1cm] {$\mathcal{D}^{(1)}_1$};
        \node (in10) [io2, below of=in11, yshift=2cm] {\rotatebox{90}{...}};
        \node (in12) [io1, right of=in1, yshift=-1cm, xshift=1cm] {$\mathcal{D}^{(J)}_1$};
        \node (in111) [io1, fill=cyan!80!violet!10, right of=in11, xshift=1cm] {$\mathcal{GP}\left(\mathcal{D}^{(1)}_1\right)$};
        \node (in110) [io2, below of=in111, yshift=2cm] {\rotatebox{90}{...}};
        \node (in121) [io1, fill=cyan!80!violet!10, right of=in12, xshift=1cm] {$\mathcal{GP} \left(\mathcal{D}^{(J)}_1\right)$};
        \node (in13) [io1, fill=orange!60!yellow!10, right of=in111, xshift=1cm, yshift=-1cm] {RCM 1 BCM};
        \node (in130) [io2, below of=in13, yshift=1.75cm] {\rotatebox{90}{...}};

        \node (in2) [io1, right of=start, yshift=-1.25cm, xshift=1cm] {RCM $R$ ($\mathcal{D}_R$)};
        \node (in21) [io2, right of=in2, xshift=1cm] {...};
        \node (in22) [io2, right of=in21, xshift=1cm] {...};
        \node (in23) [io1, fill=orange!60!yellow!10, right of=in22, xshift=1cm] {RCM $R$ BCM};

        \node (in4) [io1, fill=violet!50!pink!10, right of=in13, xshift=1cm, yshift=-1.25cm] {CORDEX MoE};
        
        \draw[arrow](start) -- (in1);
       
        \draw[arrow](in1) -- (in11);
        \draw[arrow](in1) -- (in12);
        \draw[arrow](in11) -- (in111);
        \draw[arrow](in12) -- (in121);
        \draw[arrow](in121) -- (in13);
        \draw[arrow](in111) -- (in13);

        \draw[arrow](start) -- (in2);
        \draw[arrow](in2) -- (in21);
        \draw[arrow](in21) -- (in22);
        \draw[arrow](in22) -- (in23);
        
        \draw[arrow](in13) -- (in4) node[midway, above right, xshift=-0.3cm] {\small $w_1$};
        \draw[arrow](in23) -- (in4) node[midway, below right, xshift=-0.3cm] {\small $w_R$};
    \end{tikzpicture}
    \end{adjustbox}
%TC:endignore
    \caption[Diagram of ensemble learning method]{Ensemble learning method. For a given emission scenario and climatological period, each of the $R$ RCMs outputs from CORDEX-WAS are split into $J$ manageable spatiotemporal domains, $\mathcal{D}^{(j)}_r$ for the $j$\textsuperscript{th} domain and $r$\textsuperscript{th} RCM. A GP is fit to each domain and then combined using a BCM. The outputs of each RCM BCM are then combined using a weighted mixture model or mixture-of-experts (MoE) with weights $\{w_r\}^R_{r=1}$.}
    \label{fig:bcm_diagram}
\end{figure}

\subsection{RCM surrogate models}
\label{sec:rcm_surrogates}

\subsubsection{Gaussian processes}
\label{sec:gp}

To create the surrogate models for each RCM, we consider a regression problem with a dataset of $N$ paired inputs $\bm{X} := \{\bm{x}_i\}_{i=1}^N$ and noisy observations $\bm{Y} := \{y_i\}_{i=1}^N$, where $i$ denotes the $i$\textsuperscript{th} data pair. Each input $\bm{x}_i$ is a three-dimensional vector of $(\mathrm{month}, \mathrm{latitude}, \mathrm{longitude})$. Each observation $y_i$ is a scalar, corresponding to a prediction of precipitation by an RCM.

The problem is modelled as $y_i = f(\bm{x}_i)+ \epsilon_i$, where we assume that the noise $\epsilon_i \sim \mathcal{N}(\bm{0},\sigma_\mathrm{n}^2)$ is independently and identically distributed. The function $f$ can be represented with a Gaussian process (GP) \citep{rasmussen2006gaussian}. GPs are powerful non-parametric Bayesian models. A GP is  as a distribution over functions for which every finite set of outputs is jointly Gaussian distributed. GPs are fully specified by a mean function $\mu(\cdot)$ and a kernel function $k(\cdot, \cdot)$, which depend on hyperparameters $\theta_\mu$ and $ \theta_k$, jointly referred to as $\theta$. If we assume $f$ to be a GP, i.e. $f \sim \mathcal{GP}(\mu(\cdot), k(\cdot, \cdot))$, it follows that $f(\bm{X}) \sim \mathcal{N}(\bm{\mu}_{\bm{X}},\mathbf{K}_{\bm{X}} + \sigma_\mathrm{n}^2 \mathbf{I})$ where $(\bm{\mu}_{\bm{X}})_i := \mu(\bm{x}_i)$ is a $N$-dimensional vector and $(\mathbf{K}_{\bm{X}})_{ij} = k(\bm{x}_i, \bm{x}_j)$ is an $N \times N$ covariance matrix. 

A posterior GP conditioned on the dataset $\mathcal{D} = (\bm{X}, \bm{Y})$ can be inferred. This joint probability distribution of precipitation is trained using all data at every month and location of the RCM simultaneously. The GP is then trained by learning the optimal hyperparameters from the data. This optimisation is performed by maximising the log-marginal likelihood of the model,
\begin{equation}
\hat\theta = \mathop{\mathrm{argmax}}_\theta [\log p(\bm{Y} \vert \bm{X}, \theta)] = \mathop{\mathrm{argmax}}_\theta \left[ \log \mathcal{N} \left(\bm{Y}|\bm{\mu}_{\bm{X}}(\theta_\mu), \mathbf{K}_{\bm{X}}(\theta_K) + \sigma_\mathrm{n}^{2}\mathbf{I}\right) \right].
\end{equation}

Finally, the GP can be sampled repeatedly at multiple arbitrary inputs. For a given input at an arbitrary month, latitude, and longitude, i.e. a test point $\bm{x}_*$ not necessarily seen in the training data, the mean and variance of the posterior predictive distribution of $f$ are given by
\begin{equation}
\mathop{\mathbb{E}}_{f \vert \mathcal{D}}[f(\bm{x}_*)] = \mu(\bm{x}_*) +\bm{K}_*^T (\mathbf{K}_{\bm{X}}+\sigma_\mathrm{n}^2 \mathbf{I})^{-1}({\bm{Y}}-\bm{\mu}_{\bm{X}}), 
\end{equation}
\begin{equation}
\mathop{\mathrm{Var}}_{f \vert \mathcal{D}}[f(\bm{x}_*)] = k(\bm{x}_*, \bm{x}_*) - \bm{K}_*^{T} (\mathbf{K}_{\bm{X}} + \sigma_\mathrm{n}^2\mathbf{I})^{-1} \bm{K}_*,
\end{equation}
where $\bm{K}_* = k(\bm{X}, \bm{x}_*)$ is an $N$-dimensional vector.

\subsubsection{Warped GPs}
In its original formulation, the posterior distribution of the GP is normal. However, monthly precipitation $y$ is generally log-normally distributed \citep{tazi2024downscaling}. Warped GPs handle non-normality in the data by transforming the target variables and GP outputs via a warping function \citep{snelson2003warped}. For this problem, we use a Box-Cox function $g_\lambda(\cdot)$ as the warping function. The warped outputs $\tilde{y}$ are then given by
\begin{equation}
\tilde{y} = g_\lambda(y)= 
    \begin{cases}
    \frac{y^{\lambda} - 1 }{\lambda} & \text{for} \quad \lambda \neq 0, \\
    \log y  & \text{for} \quad \lambda = 0,
    \end{cases}       
\end{equation}
where precipitation $y > 0$ is assumed to be positive and $\lambda$ is the scaling factor. The scaling factor is optimised to make the distribution of $g_\lambda(y)$ as Gaussian as possible by maximising the log-likelihood of the model parameters arising from placing a Gaussian distribution over the transformed observations \citep{box1964analysis}. In this study, we use a different optimised scaling factor $\lambda_{r}$ for each RCM $r$. 

Finally, to recover the skewness and more complex distributional characteristics of the RCM data, the corresponding inverse transform is applied to the GP outputs. Henceforth, we will use $f(\cdot)$ to denote the function that models the \emph{warped} output $\tilde{y}$. We will denote the unwarped function, which may have heavy tails in its distribution, as $f_\lambda(\cdot) := g^{-1}_\lambda (f(\cdot))$. Details on the distribution of the unwarped precipitation can be found in \cref{apdx:moe-dist}.

\subsubsection{Chained GPs}

For this application, the variance of precipitation $y$ should be considered as intrinsic climatological variance rather than noise. The GP output should therefore be heterogeneous in its variance, i.e. heteroskedastic. To address these requirements, we use chained GPs \citep{saul2016chained} to model the variance and mean as latent functions such that
\begin{equation}
p(\tilde{y}_i \vert f_1, f_2, \bm{x}_i) = \mathcal{N}\Big(\tilde{y}_i \, \Big\vert \, f_1(\bm{x}_i), \alpha\big(f_2(\bm{x}_i)\big)\Big),
\end{equation}
where $f_1 \sim \mathcal{GP}(0, k_1(\cdot, \cdot))$,  $f_2 \sim \mathcal{GP} (0, k_1(\cdot, \cdot))$ and $\alpha$ is a function used to map the outputs of $f_2$ to positive values. This mapping is required as the output of $\alpha$ represents the variance of a Gaussian distribution. Here, we apply an exponential function $\alpha(f_2(\bm{x}_i)) = e^{f_2(\bm{x}_i)}$, but other functions, such as softplus, could be used. For our work, we use a shortcut to calculate the variance at each point without using further approximations, such as inducing points \citep{titsias2009variational}, by exploiting the gridded nature of our data.

\subsubsection{Bayesian committee machines}
\label{sec:bcm}

Due to GP's poor scalability to large datasets, we use robust Bayesian committee machines (BCMs) as a way of combining multiple GPs and speeding up inference \citep{tresp2000bayesian, deisenroth2015distributed, cohen2020healing}. To do this, the training dataset is partitioned into $J$ subsets of size $M$ (so $N=JM$) where $M \ll N$. We define these subsets or domains as $\mathcal{D}^{(j)} = (\bm{X}^{(j)}, \bm{Y}^{(j)})$ for $j=1, \dots, J$.  GPs are then used to model each domain. The robust BCM GPs share kernel hyperparameters, stopping individual experts from overfitting their local subset of data. Assuming conditional independence between domains $\mathcal{D}^{(j)}$ and by repeated application of Bayes' theorem, the following predictive distribution is obtained by combining the separate GP models:
\begin{equation}
p(f(\bm{x}_*) \vert \mathcal{D}) = \frac{\prod_{j=1}^{J}  p_{j}^{\beta_{j}(\bm{x}_*)}(f(\bm{x}_*) \vert \mathcal{D}^{(j)})}{p^{-1+\sum_j\beta_{j}(\bm{x}_*)}(f(\bm{x}_*))},
\end{equation}
where $p(f(\bm{x}_*))$ is the prior distribution of the GP evaluated at test point $\bm{x}_*$ and $\beta_{j}$ controls the contribution of expert $j$ (see \cref{app:softmax-variance}). %This distribution is inherently Gaussian as the product of Gaussian distributions is also Gaussian.
The robust BCM’s predictive mean and precision (inverse of the variance) are given by
\begin{equation}
\mu_*^\text{rbcm} := \mathop{\mathbb{E}}_{f \vert \mathcal{D}^{(1)} \cdots \mathcal{D}^{(j)}} \left[ f(\bm{x}_*) \right] = (\sigma^{\text{rbcm}}_*)^{2} \sum^{J}_{j=1} \beta_{j}(\bm{x}_*)(\sigma^{j}_*)^{-2} \mu^j_*,
\end{equation}
\begin{equation}
(\sigma_*^{\text{rbcm}})^{-2} := \mathop{\mathrm{Var}}_{f \vert \mathcal{D}^{(1)} \cdots \mathcal{D}^{(j)}} \left[ f(\bm{x}_*) \right]^{-1} = (1-M) (\sigma^\text{prior}_*)^{-2} + \sum^{J}_{j=1} \beta_{j}(\bm{x}_*)(\sigma^j_*)^{-2},
\end{equation}
respectively, where $\mu^j_*$ and $(\sigma^j_*)^{-2}$ are the mean and precision of the predictive posterior given domain $\mathcal{D}^{(j)}$, and $(\sigma^\text{prior}_{*})^{-2}$ is the prior precision of $p(f_*)$. This setup conserves the properties of an exact GP while improving scalability, moving from $\mathcal{O}(N^3)$ for training and $\mathcal{O}(N^2)$ for prediction to $\mathcal{O}(JM^3)$ and $\mathcal{O}(JM^2)$ respectively \citep{deisenroth2015distributed}. For our work, the domains are divided temporally, such that each domain contains precipitation for a particular month over the whole spatial area. 

\begin{figure}[ht]
    \centering
    \includegraphics[width=\textwidth]{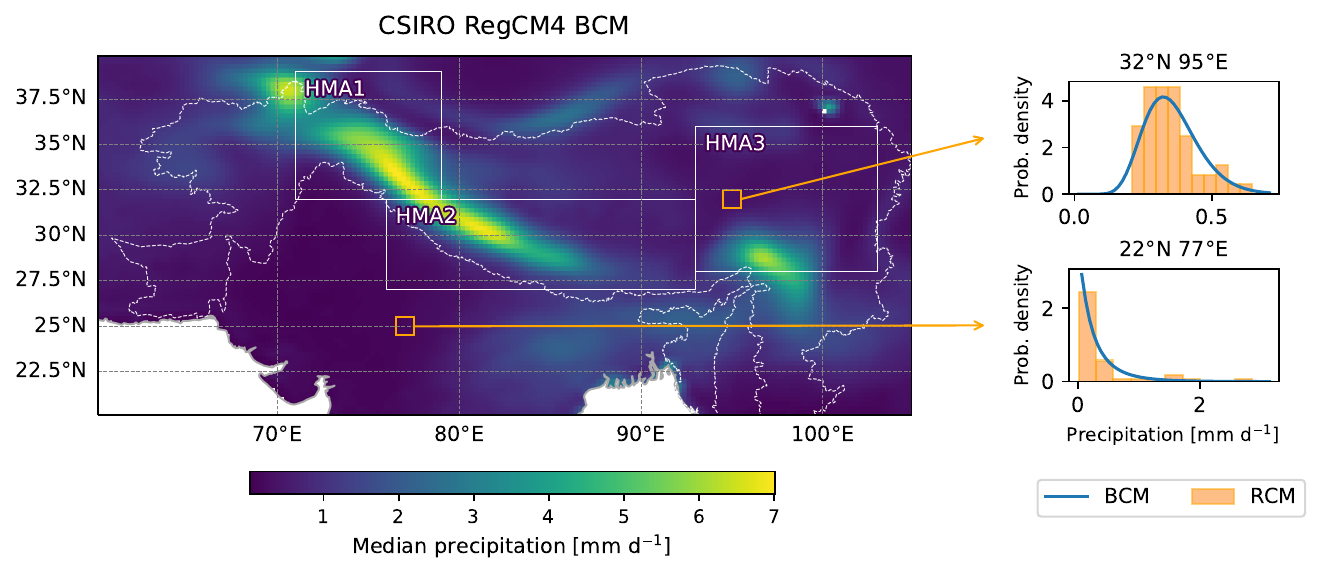}
    \caption[CSIRO RegCM4 BCM distributions between 1951 and 2005 for February]{CSIRO RegCM4 BCM distributions between 1951 and 2005 for the month of February. The figure shows the median of the surrogate distributions over HMA (left). The surrogate distributions (blue line) are compared with the empirical histograms (orange bars) for two example locations (top and bottom right). The full surrogate distributions are faithful to the original RCM distributions, which differ significantly depending on time and location.}
    \label{fig:bcm_output}
\end{figure}

With these adaptations, GPs can be used to interpolate large empirical distributions. To compare the RCM surrogates with the APHRODITE dataset introduced in \cref{sec:aphro}, we evaluate the surrogate GPs at the APHRODITE grid points. In \cref{fig:bcm_output}, we illustrate the typical surrogate output with CSIRO RegCM4 for February. The surrogate distributions %(blue)
closely match the empirical RCM histograms,
%(orange)
which can substantially differ with time and location.

% \subsubsection{Practical distribution calculation}

% Each CORDEX RCM currently sits on a shared grid
% \begin{enumerate}
%     \item Aggregate data to months and locations on grid
%     \item Apply Box-Cox transformation to the data
%     \item Calculate the mean and the variance
%     \item Z-score the mean and variance
%     \item Divide the data by months resulting in 40 by 90 grids (3600 points)
%     \item Fit 12 GP experts to model the mean and another 12 to model the variance
%     \item Inverse transform z-scoring
% \end{enumerate}

\subsection{Surrogate aggregation}
\label{sec:across_rcms}

\subsubsection{Mixture of experts}
\label{sec:moe}

The RCM surrogates are aggregated via the weighted sum of their posterior probability distributions. We refer to this weighted mixture model as a mixture of experts (MoE). The resulting distribution for a given test point $\bm{x}_*$ is
\begin{equation}
p_\mathrm{MoE}(f_{\bm{\lambda}}(\bm{x}_*) \vert \bm{w}, \mathcal{D}) = \sum_{r=1}^R w_r p_r(f_{\lambda_r}(\bm{x}_*) \vert \mathcal{D}_r).
\end{equation}
Here, $\mathcal{D}$ is the dataset for all RCMs, and $\mathcal{D}_r$ is the dataset for the $r$\textsuperscript{th} RCM for $r=1, \dots, R$. The distribution $p_r(f_{\lambda_r}(\bm{x}_*) \vert \mathcal{D}_r)$ is the \emph{unwarped} $r$\textsuperscript{th} RCM surrogate distribution conditioned on $\mathcal{D}_r$. It is \emph{not} Gaussian and depends on the warping parameter $\lambda_r$. The scalar $w_r = w_r(\bm{x}_*)$  is the weight of the $r$\textsuperscript{th} RCM at $\bm{x}_*$. It is constrained to be $\sum_{r=1}^R w_r = 1$ and $w_r>0$.

\subsubsection{Weights}

We would like to choose weights $\bm{w}$ such that the MoE accurately predicts future and past precipitation distributions. Without prior knowledge, the simplest weighting would be $w_r = 1/R$, i.e. an equally-weighted (EW) mixture of the RCM surrogates, not assigning more importance to any particular RCM. Instead, we inform the weights through a reference dataset (APHRODITE; see \cref{sec:aphro}), having previously downscaled RCM predictions onto the same grid using the surrogate models, as shown in \cref{sec:gp,fig:bcm_output}.

A naive approach would be to directly learn the weights that maximise the likelihood of the parameters (APHRODITE). However, there are two pitfalls to this method. First, the number of effective parameters ($R-1$ =12), compared to the number of datapoints at each location and time (${\sim}30$), results in overfitting the $R$ model weights (not shown). Second, direct maximum likelihood optimisation could upweigh RCM surrogates that individually predict precipitation poorly, but when summed together best approximate APHRODITE. We want to ensure that RCM surrogates that contribute significantly to the aggregate are independently close fits to APHRODITE. This condition prevents error compensation in the MoE.

Thus, we inform the weights using a measure of distance between the distribution of each RCM surrogate and that of APHRODITE. We choose the Wasserstein distance \citep{panaretos2019statistical}, which originates from optimal transport theory and represents the minimum cost of moving the probability density of one distribution into the shape of another. For two empirical distributions $P$ and $Q$, the Wasserstein distance $W$ is given by:
\begin{equation}
    W (P, Q) = \left( \frac{1}{N} \sum^N_{i=1} \Vert X_{i} -  Y_{i} \Vert^p \right)^{1/p},
\end{equation}
where $X_i$ and $Y_{i}$ are the samples of $P$ and $Q$, respectively, and $p$ is the moment of the distribution we are interested in. Here, we use the first moment $p=1$, which represents the shift of the distribution mean. We calculate the Wasserstein distances with respect to a reference historical period for each APHRODITE grid point and month. Previous research has shown that evaluating the shapes of the distribution rather than the differences in the absolute precipitation outputs better distinguishes how well the models represent precipitation \citep{martinez2021climate}. We therefore scale the 95\textsuperscript{th} percentile of the BCM outputs to match those of APHRODITE and thus avoid comparing RCMs based on systematic biases in their precipitation output.

We then introduce the statistical temperature $T$, which interpolates between choosing the most accurate RCM ($T = 0$) and equally weighting the RCMs ($T \rightarrow \infty$). Using this concept, the weights are parametrised as
    \begin{align}
    w_r = \frac{e^{-h(W_r)/T}}{\sum_{r'} e^{-h(W_{r'})/T}},
    \end{align}
where $h(\cdot)$ is some monotonically increasing function. The statistical temperature $T$ is then optimised to maximise the likelihood on the APHRODITE dataset. We chose $h(\cdot) = \ln(\cdot)$, which reduces the variation between the Wasserstein distances. This choice meets our criteria for the weights to vary smoothly in space and to not choose a single RCM. To further stabilise optimisation, we maximised the likelihood around a $5 \times 5$ grid surrounding each location. The RCM surrogates with the smallest Wasserstein distances and the maximum MoE weights over HMA are shown in \cref{sec:best_rcm}.

% \subsubsection{Practical distribution calculation}

% (This section will most likely go to the Appendix)
% \begin{enumerate}
   
%     \item We then sample the resulting heavy tailed distributions (100 times?)
    
%     \item The model outputs are retransformed into [mm/day] using the mean and the variance.
    
%     \item Scale APHRODITE and each BCM distribution by their own95\textsuperscript{th} percentile  
    
%     \item Calculate the empirical Wasserstein distance

%     \item Weights are calculated through some normalisation

%     \item Weights are applied to the future prediction distribution (these need to be scaled with the same 95\textsuperscript{th} percentile value as earlier)
% \end{enumerate}

\section{MoE validation}
\label{sec:hist_validation}

% With the Wasserstein distances calculated, the weights for the MoE are learnt. The performance of the MoE can then be evaluated against an equally-weighted mixture of the surrogate's samples (EW). This time, 
To evaluate the performance of the MoE approach relative to the EW, the historical experiment is divided into two sets: a training period (1951--1980) and a validation period (1981--2005). We use the continuous rank probability score \citep[CRPS,][]{hersbach2000decomposition} to measure the distance between the predicted (MoE or EW) and observed (APHRODITE) cumulative distribution functions (CDF) $F$. The CRPS is defined as the integral of the squared difference between the predicted CDF $F(y)$ and the observed CDF $F_\text{obs}(y)$ over the range of possible values $y$: % (Eq. \ref{eq:CRPS}).
% CRPS
\begin{equation}
    \text{CRPS} = \int_{-\infty}^{\infty} ({F}(y) - F_\text{obs}(y))^2 dy.
   \label{eq:CRPS}
\end{equation}
Smaller scores therefore imply more skill. 

\begin{figure}
\center
\includegraphics[width=\textwidth]{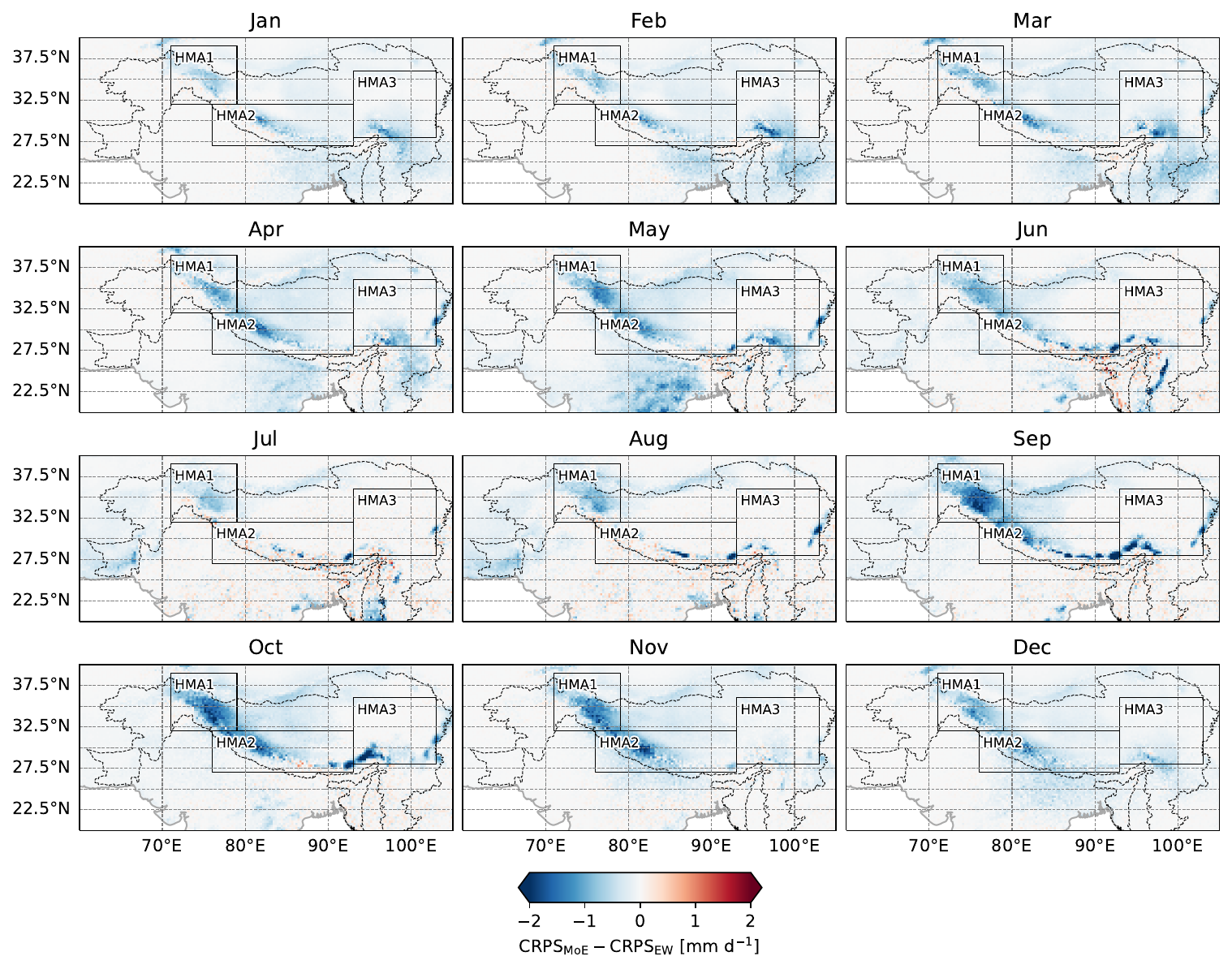}
\caption[MoE and EW CRPS differences over the validation period]{MoE and EW CRPS differences over the held-out validation period (1981--2005). The MoE and EW outputs are generated from $10^5$ RCM surrogate samples for each month and location. Negative values (blue) imply MoE matches APHRODITE more closely while positive values (red) represent times and locations where the EW performs better. The MoE outperforms the EW for each month, in particular over the Karakoram and Himalayan arc.}
\label{fig:crps_diff}
\end{figure}

The spatiotemporal distribution of the differences between the MoE and EW CRPS are presented in \cref{fig:crps_diff}. The MoE CRPS values are in most cases smaller than the EW CRPS values, with the MoE approach yielding an average improvement of 31\% over EW. The greatest improvements are over the Karakoram mountains and Himalayas, with an annual average difference of  \SI{-0.34}{\mm \per \day} for the West Himalaya and Karakoram (HMA1), \SI{-0.24}{\mm \per \day} for the Central and East Himalaya (HMA2), and \SI{-0.14}{\mm \per \day} for the Hengduan Shan and Southeast Tibet (HMA3).
% ; while EW is sometimes more advantageous over the south Himalaya during the summer months. 

The differences between MoE and EW can be explained by the relative contribution of RCM members to the ensemble distributions. For example, over certain areas along the central-eastern Tibetan plateau the MoE method assigns dominant weight to one or two RCM surrogates (maximum weight $\geq$ 0.7, \cref{fig:max_weights}). On the other hand, EW dilutes the skill by not down-weighting members with poorer performance, thus broadening the predictive distribution and inflating CRPS.

In contrast, areas where no significant change in scores is observed tend to correspond to locations where the MoE learns close to equal weighting (see \cref{fig:max_weights}). This occurs in the south Himalaya during the summer monsoon, where the spread in Wasserstein distance between experts collapses and the learned weights revert towards 1/13. Here, MoE gains largely disappear. Indeed, EW sometimes shows marginally superior performance. Therefore, the spatial pattern of CRPS improvement mirrors the pattern of weight concentration: where model skill heterogeneity is large, MoE expert selection yields significant gains; where expert skill converges, EW averaging suffices. The MoE also improves over taking any individual expert, i.e. RCM surrogate, by 254\% as illustrated in \cref{fig:moe_vs_rcms,fig:scorecard}.

% The Brier score (BS) measures the mean squared error between $M$ pairs of exceedance probabilities $\pi_m$ and binary observations $r^\text{obs}_m$ for a specified precipitation event (i.e., precipitation threshold) (Eq. \ref{eq:BS}). 

% % BS
% \begin{equation}
%     \text{BS} = \frac{1}{M}\sum_{m=1}^{M}(\pi_m - r_{m}^\text{obs})^2.
%    \label{eq:BS}
% \end{equation}

% The Kolmogorov-Smirnov statistic quantifies the distance between two empirical cumulative distribution functions. [MORE.]

% The Anderson-Darling test... 

% KL divergence

% Timeseries vs climatology

% \subsubsection{Plots}
% Reliability diagrams

% Outputs of distributions vs taking mean and variance

\section{Refined climatologies}
\label{sec:refined_clima}

To generate refined historical and future climatologies, we optimise the weights over the entire historical period (1951--2005), maximising the data available for training. We then use these weights to generate MoE ensemble predictions for the same climate scenarios and time periods as \citet{sanjay2017downscaled}: i) historical (1976--2005), ii) RCP4.5 near (2036--2065) and future (2066--2095), and iii) RCP8.5 near and far future. In the following analysis, we focus on historical and far-future periods to generate projected precipitation trends relative to a historical baseline. The near-future projections show similar trends to the far future, but with less pronounced changes, and are included in \cref{app:near_future}.

\subsection{Historical predictions}
\label{sec:historical_preds}

\begin{figure}[ht]
\center
\includegraphics[width=\textwidth]{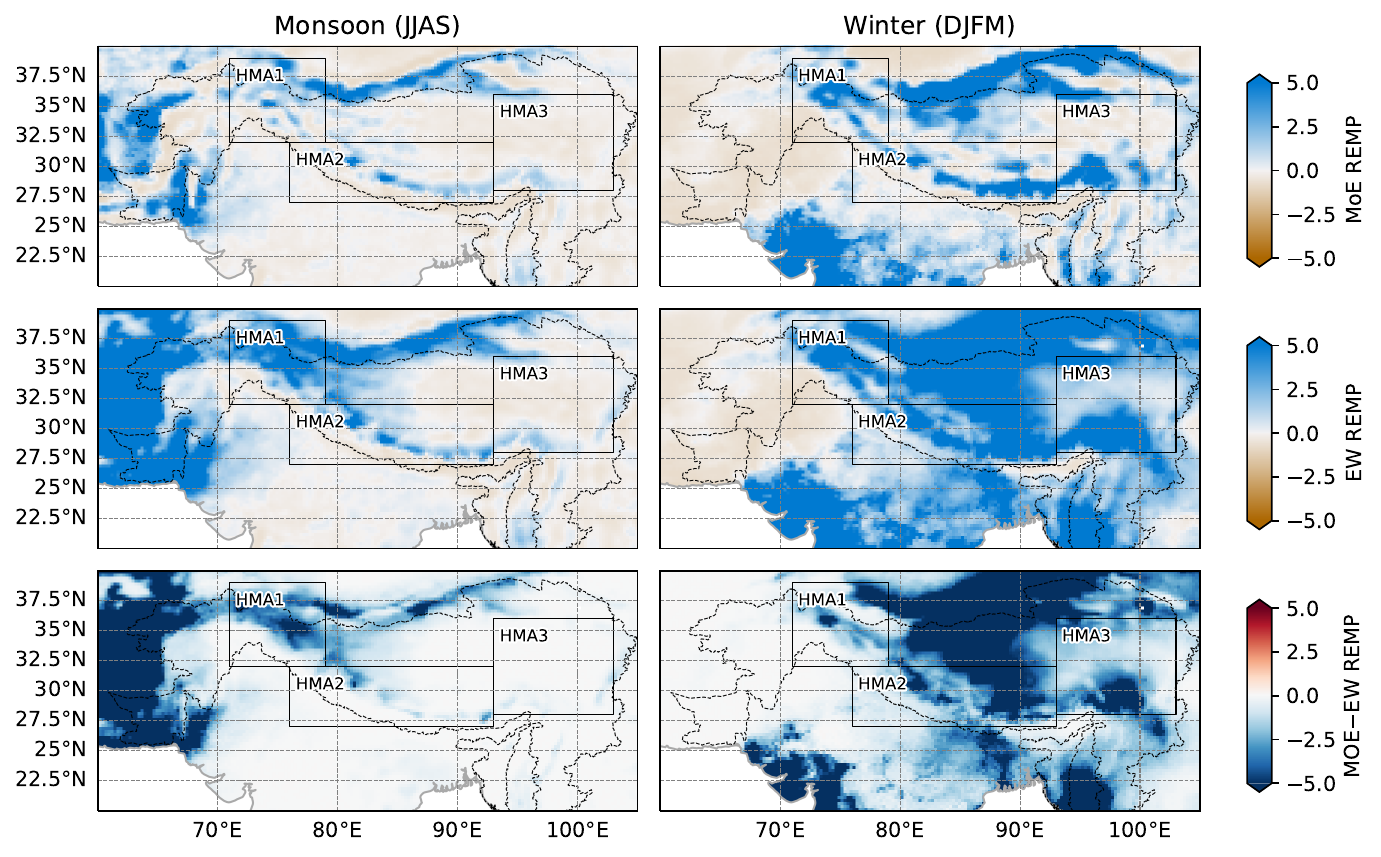}
\caption[Historical MoE and EW relative error with respect to APHRODITE]{Historical MoE and EW relative error for median precipitation (REMP) with respect to APHRODITE over HMA. The MoE REMP (top), the EW REMP (middle), and the REMP difference between MoE and EW (bottom) are plotted for the historical reference period (1976--2005) for the summer monsoon (left) and winter (right).  The MoE and EW outputs are generated from $10^5$ RCM surrogate samples for each month and location. The MoE makes large improvements over the EW, in particular, over locations where precipitation is overestimated by the EW.}
\label{fig:moe_hist}
\end{figure} 

\cref{fig:moe_hist} plots the relative error for median precipitation (REMP) of the MoE and EW with respect to APHRODITE for the historical period. For the monsoon season, the MoE and EW both under- and overestimate precipitation for the Himalaya and Karakoram (HMA1 and HMA2). More specifically, the models predict a wet bias for the Karakoram, inner Himalayas, and western Tibetan Plateau and a dry bias for the outer Himalayas.
Precipitation over the eastern Tibetan Plateau, Hengduan Shan, and Southeast Tibet (HMA3) is generally underestimated by both models. For winter, the MoE again under- and overestimates precipitation across the three studied subregions, but with larger wet biases, while the EW  largely overestimates precipitation across all three areas during winter.

The third row of \cref{fig:moe_hist} plots the difference between the MoE and EW REMP. The MoE makes large improvements over the EW, particularly in locations where the EW overestimates precipitation. This corresponds to a difference in REMP of 2.02 for the West Himalaya and Karakoram (HMA1), 0.32 for the Central and East Himalaya (HMA2), and 0.08 for the Hengduan Shan and Southeast Tibet (HMA3) over the summer. During winter, larger improvements are obtained with REMP differences of 1.58, 3.02, and 2.33 for the West Himalaya and Karakoram (HMA1), the Central and East Himalaya (HMA2), and the Hengduan Shan and Southeast Tibet (HMA3), respectively.
In summary, the MoE predicts smaller median precipitation values compared to the EW, more in line with APHRODITE. However, MoE predictions are still subject to large errors, and bias correction would still be needed before use. For completeness, we include the historical predictions biases in \unit{\mm \per \day} in \ref{app:hist_pred_abs}.

\subsection{Future projections}
\label{sec:future_preds}

\begin{figure}[ht]
\center
\includegraphics[width=\textwidth]{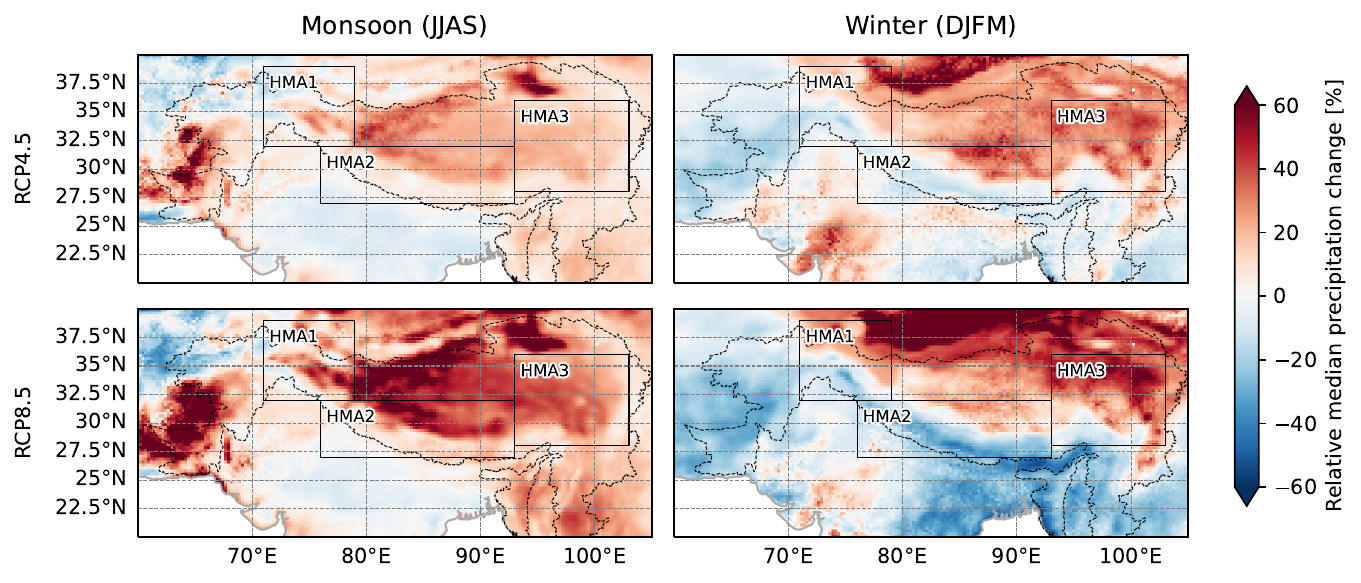}
\caption[Relative changes between far-future and historical MoE precipitation]{Relative changes between far-future (2066--2095) and historical (1976--2005) MoE precipitation. The plot shows the median changes for RCP4.5 (top) and RCP8.5 (bottom) during the summer monsoon (left) and winter (right) for the far-future across HMA.  The MoE output is generated from $10^5$ RCM surrogate samples for each month and location. The MoE predicts large departures from the precipitation predicted for the historical reference period.}
\label{fig:moe_preds}
\end{figure}

\cref{fig:moe_preds} shows far-future MoE projections with respect to the historical period. For RCP4.5 and RCP8.5, the monsoon season is projected to see an increase in median precipitation over most of HMA, with the exception of the north Hindu Kush mountains. The greatest changes occur over Kunlun mountains and Southeast Tibet. More specifically, the MoE predicts an average increase of 15\% for the West Himalaya and Karakoram (HMA1), 23\% for the Central and East Himalaya (HMA2), and 27\% for the Hengduan Shan and Southeast Tibet (HMA3) for RCP8.5. Compared to RCP4.5, RCP8.5 presents a more pronounced wetting trend over the study area.

For winter, the MoE projects an increase in median precipitation over the north and northeast HKH, with the largest changes occurring over the Quilian Shan and East Kunlun mountains. The Hindu Kush mountains and the outer Himalayas show an overall decrease in precipitation for this season. These trends lead to mixed predictions for the West Himalaya and Karakoram (HMA1) and the Central and East Himalaya (HMA2), with average relative changes of 15\% and -7\% for RCP8.5, respectively. Winter precipitation over the Hengduan Shan and Southeast Tibet (HMA3) is expected to increase with an average relative change of 27\%. Compared to RCP4.5, RCP8.5 generally presents more pronounced wetting and drying trends during winter.

\begin{figure}[ht]
\center
\includegraphics[width=\textwidth]{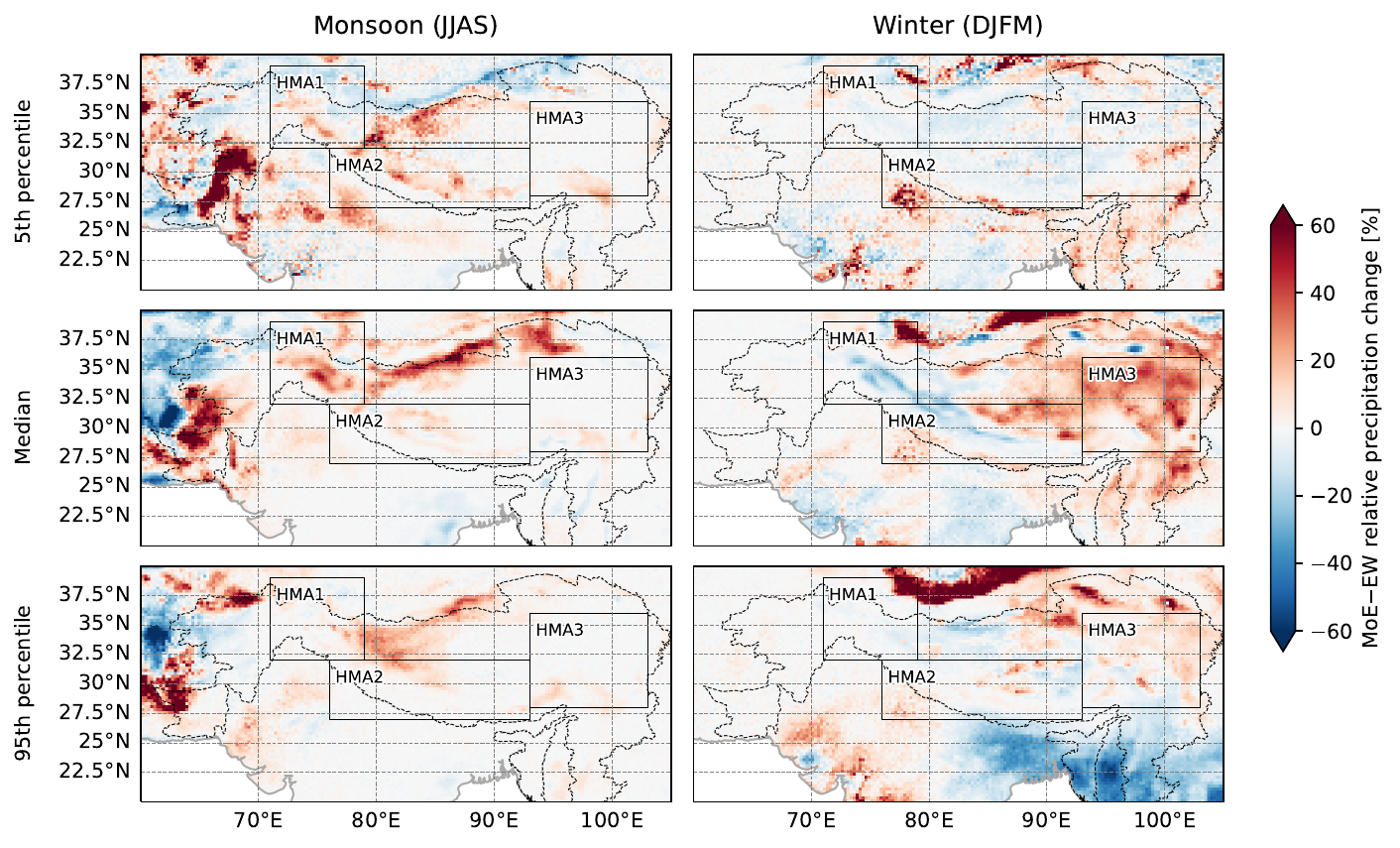}
\caption[MoE and EW relative prediction differences for the far future under RCP8.5]{MoE and EW relative prediction differences for the far future (2066--2095) under RCP8.5 across HMA with respect to their historical reference predictions. The plot shows the differences between the predicted relative changes for the 5\textsuperscript{th} percentile (bottom), median (middle), and 95\textsuperscript{th} percentile (top) during the summer monsoon (left) and winter (right). The MoE and EW outputs are generated from $10^5$ RCM surrogate samples for each month and location. Although for many locations the MoE and EW projections are similar, there are significant difference in predictions during both the summer monsoon and winter over key mountain ranges.}
\label{fig:ff_pred_diffs}
\end{figure}

Finally, the relative changes in precipitation from the MoE can be compared with those from the EW. \cref{fig:ff_pred_diffs} plots the difference between the MoE and EW relative changes for the far-future RCP8.5 scenario with respect to their historical reference predictions. The plot shows that for large areas, including the Tibetan Plateau over summer and the Hindu Kush, the relative precipitation changes predicted by the MoE and the EW stay within ±10\% across different percentiles. However, many locations also present significant changes. 

During the summer monsoon, positive differences in predicted precipitation changes (i.e., MoE increase relative to EW) are observed over the West Himalaya, Karakoram and Kunlun mountains with median and 95\textsuperscript{th} percentile differences reaching maximum values of 54\% and 32\% for HMA1. This suggests that the probability of events associated with very high precipitation rates, such as floods and landslides, could be higher compared to previous estimates. During winter, the MoE predicts a greater decrease in projected precipitation over the inner Himalayas and Karakoram, with median differences achieving minimum values of -31\% for HMA1 and -27\% for HMA2. This could signify a decrease in the contribution of winter precipitation to solid water resources. At the same time, large positive differences are observed over the Tibetan Plateau and Southeast Tibet, with maximum differences in median values of 51\% for HMA2 and 62\% for HMA3.

\section{Discussion}
\label{chap4-sec:discussion}

%\subsection{Contributions}
This paper explored probabilistic ensemble learning to aggregate RCMs.  Using a MoE model, precipitation predictions were improved compared to an EW approach. Unlike other ensembling techniques, this method does not narrow the non-discountable envelope of climate change, but refines it. With GPs as its foundational building blocks, the MoE should work equally well with RCMs on different grids and can be straightforwardly applied to any kind of model ensemble. However, this study does present several limitations.

%\subsection{Limitations}
First, we assume that the driving variables from the driving GCMs are sufficiently well-modelled to dynamically downscale precipitation. In reality, CMIP5 GCMs present large differences in their predictions over HMA \citep{panday2015cmip, palazzi2015precipitation}. Moreover, recent studies incorporating more realistic aerosol scenarios further amplify future precipitation uncertainty \citep{jiang2023precipitation, wilcox2020accelerated}. This poor representation of precipitation in RCMs can be observed in this study through the optimal weights chosen by the MoE: when no model in the ensemble performs particularly well, the MoE chooses weights close to the EW. This is the case for the Hindu Kush during winter and for the Tibetan Plateau during summer (cf. \cref{fig:moe_hist} and \cref{fig:max_weights}). This behaviour, in turn, implies poor representation of the two main precipitation drivers for HMA: the East Asian summer monsoon and western disturbances. Further work could use this framework to explain performance differences across RCMs by analysing other output variables, such as pressure or humidity fields. 

Second, the methods used to create the RCM surrogates have several limitations. By applying GPs, we assume that the RCM outputs are unimodal. In reality, this is not always the case, and we could be missing out on inter-annual variability. The degree to which the individual GP samples are consistent in time and space is also uncertain.. Using a probabilistic surrogate for the RCM means that future precipitation scenarios can be sampled thanks to the correlations learnt between different locations and times. However,  by aggregating data over time, the RCM surrogate, and by extension the MoE, could plausibly sample two high precipitation values for neighbouring locations or months, although the joint event is statistically unlikely. Reintroducing some of the information from the RCM time series would be needed to ensure further consistency.

Third, the weighting function used to aggregate the RCMs, although well motivated, has an arbitrary form. There is significant room to refine the approach through more case studies. 
%%%%% Cumulative dry and wet days
Finally, monthly precipitation is not the only variable linked to water security over HMA. To get a more holistic view of future changes, and thus the risk of extreme events such as floods and droughts, the experiments in this paper could be repeated with variables such as temperature or the number of cumulative dry and wet days. Extreme value analysis could also be applied to calculate updated rates of return for single or compounding hydrological extremes.

%%%% Aggregating RCMs from different with large difference in historical data 

%\subsection{Further work}

\section{Conclusion}
\label{sec:conclusion}

This paper investigated a mixture of experts (MoE) approach to aggregate an ensemble of RCMs. This ensemble learning method is generalisable to other climate variables and model ensembles and could be applied to model aggregation problems in other fields. The MoE was applied to the CORDEX-WAS ensemble over HMA and was found to be more accurate than the equally-weighted average (EW) or any single ensemble member for a held-out validation period. %For large areas, including the Tibetan Plateau over summer and the Hindu Kush in winter, the MoE and EW precipitation projections stay with ±10\% across their predicted distributions for the RCP8.5 and RCP4.5 far future (2066--2095). However, 
Compared to the EW, the MoE projects wetter summers but drier winters over the western Himalayas and Karakoram and wetter winters over the Tibetan Plateau, Hengduan Shan, and South East Tibet for both the near and far future under RCP4.5 and RCP8.5.

%TC:ignore

\section*{Software and data accessibility}
The code and data for this paper are available on Zenodo (\url{https://doi.org/10.5281/zenodo.14907030}, \url{https://doi.org/10.5281/zenodo.14837271}). The CORDEX-WAS data are available through the Earth System Federation Grid nodes \url{https://esgf-metagrid.cloud.dkrz.de} and the APHRODITE data through the APHRODITE's Water Resources website \url{http://aphrodite.st.hirosaki-u.ac.jp/download/}.

\section*{Acknowledgements}
The authors thank Andrew Orr and Scott J. Hosking for their thoughtful feedback on the manuscript. This work was supported by the Engineering and Physical Sciences Research Council through the AI for Environmental Risk (AI4ER) CDT [grant number: EP/S022961/1] for KT and MGM, and the DTP International Studentship [grant number: EP/W524475/1] for SWPK.

\bibliography{main}

\begin{thebibliography}{68}
\providecommand{\natexlab}[1]{#1}
\providecommand{\url}[1]{\texttt{#1}}
\expandafter\ifx\csname urlstyle\endcsname\relax
  \providecommand{\doi}[1]{doi: #1}\else
  \providecommand{\doi}{doi: \begingroup \urlstyle{rm}\Url}\fi

\bibitem[Amos et~al.(2022)Amos, Pinder, and Young]{amos2022identifying}
M.~Amos, T.~Pinder, and P.~Young.
\newblock Identifying latent climate signals using sparse hierarchical {Gaussian} processes.
\newblock In \emph{NeurIPS 2022 Workshop on Tackling Climate Change with Machine Learning}, online, 9 December 2022.
\newblock URL \url{https://www.climatechange.ai/papers/neurips2022/15}.
\newblock Accesssed 18 November 2024.

\bibitem[Bajracharya and Shrestha(2011)]{bajracharya2011status}
S.~R. Bajracharya and B.~R. Shrestha.
\newblock The status of glaciers in the {Hindu Kush-Himalayan} region.
\newblock International Centre for Integrated Mountain Development (ICIMOD), 2011.
\newblock URL \url{https://doi.org/10.53055/ICIMOD.551}.

\bibitem[Box and Cox(1964)]{box1964analysis}
G.~E. Box and D.~R. Cox.
\newblock An analysis of transformations.
\newblock \emph{Journal of the Royal Statistical Society: Series B (Methodological)}, 26\penalty0 (2):\penalty0 211--243, 1964.
\newblock URL \url{https://doi.org/10.1111/j.2517-6161.1964.tb00553.x}.

\bibitem[Chylek et~al.(2011)Chylek, Li, Dubey, Wang, and Lesins]{chylek2011observed}
P.~Chylek, J.~Li, M.~Dubey, M.~Wang, and G.~Lesins.
\newblock Observed and model simulated 20th century {Arctic} temperature variability: Canadian earth system model {CanESM2}.
\newblock \emph{Atmospheric Chemistry and Physics Discussions}, 11\penalty0 (8):\penalty0 22893--22907, 2011.
\newblock URL \url{https://doi.org/10.5194/acpd-11-22893-2011}.

\bibitem[Cohen et~al.(2020)Cohen, Mbuvha, Marwala, and Deisenroth]{cohen2020healing}
S.~Cohen, R.~Mbuvha, T.~Marwala, and M.~Deisenroth.
\newblock Healing products of {G}aussian process experts.
\newblock In \emph{Proceedings of the 37th International Conference on Machine Learning}, volume 119 of \emph{Proceedings of Machine Learning Research}, pages 2068--2077, online, 07 2020. PMLR.
\newblock URL \url{https://doi.org/10.48550/arXiv.2102.07106}.

\bibitem[Deisenroth and Ng(2015)]{deisenroth2015distributed}
M.~Deisenroth and J.~W. Ng.
\newblock Distributed {Gaussian} processes.
\newblock In \emph{Proceedings of the 32nd International Conference on Machine Learning}, volume~37 of \emph{Proceedings of Machine Learning Research}, pages 1481--1490, Lille, France, 07--09 Jul 2015. PMLR.
\newblock URL \url{https://proceedings.mlr.press/v37/deisenroth15.html}.
\newblock Accessed 24 January 2025.

\bibitem[Dimri(2021)]{dimri2021bias}
A.~Dimri.
\newblock Bias correction demonstration in two of the {Indian Himalayan} river basins.
\newblock \emph{Journal of Water and Climate Change}, 12\penalty0 (4):\penalty0 1297--1309, 2021.
\newblock URL \url{https://doi.org/10.2166/wcc.2020.119}.

\bibitem[Dufresne et~al.(2013)Dufresne, Foujols, Denvil, Caubel, Marti, Aumont, Balkanski, Bekki, Bellenger, Benshila, Bony, Bopp, Braconnot, Brockmann, Cadule, Cheruy, Codron, Cozic, Cugnet, de~Noblet, Duvel, Ethé, Fairhead, Fichefet, Flavoni, Friedlingstein, Grandpeix, Guez, Guilyardi, Hauglustaine, Hourdin, Idelkadi, Ghattas, Joussaume, Kageyama, Krinner, Labetoulle, Lahellec, Lefebvre, Lefevre, Levy, Li, Lloyd, Lott, Madec, Mancip, Marchand, Masson, Meurdesoif, Mignot, Musat, Parouty, Polcher, Rio, Schulz, Swingedouw, Szopa, Talandier, Terray, and Vuichard]{dufresne2013climate}
J.-L. Dufresne, M.-A. Foujols, S.~Denvil, A.~Caubel, O.~Marti, O.~Aumont, Y.~Balkanski, S.~Bekki, H.~Bellenger, R.~Benshila, S.~Bony, L.~Bopp, P.~Braconnot, P.~Brockmann, P.~Cadule, F.~Cheruy, F.~Codron, A.~Cozic, D.~Cugnet, N.~de~Noblet, J.-P. Duvel, C.~Ethé, L.~Fairhead, T.~Fichefet, S.~Flavoni, P.~Friedlingstein, J.-Y. Grandpeix, L.~Guez, E.~Guilyardi, D.~Hauglustaine, F.~Hourdin, A.~Idelkadi, J.~Ghattas, S.~Joussaume, M.~Kageyama, G.~Krinner, S.~Labetoulle, A.~Lahellec, M.-P. Lefebvre, F.~Lefevre, C.~Levy, Z.~X. Li, J.~Lloyd, F.~Lott, G.~Madec, M.~Mancip, M.~Marchand, S.~Masson, Y.~Meurdesoif, J.~Mignot, I.~Musat, S.~Parouty, J.~Polcher, C.~Rio, M.~Schulz, D.~Swingedouw, S.~Szopa, C.~Talandier, P.~Terray, and N.~V. .~N. Vuichard.
\newblock Climate change projections using the {IPSL-CM5 Earth System Model}: from {CMIP3} to {CMIP5}.
\newblock \emph{Climate dynamics}, 40:\penalty0 2123--2165, 2013.
\newblock URL \url{https://doi.org/10.1007/s00382-012-1636-1}.

\bibitem[Dunne et~al.(2012)Dunne, John, Adcroft, Griffies, Hallberg, Shevliakova, Stouffer, Cooke, Dunne, Harrison, Krasting, Malyshev, Milly, Phillipps, Sentman, Samuels, Spelman, Winton, Wittenberg, and Zadeh]{dunne2012gfdl}
J.~P. Dunne, J.~G. John, A.~J. Adcroft, S.~M. Griffies, R.~W. Hallberg, E.~Shevliakova, R.~J. Stouffer, W.~Cooke, K.~A. Dunne, M.~J. Harrison, J.~P. Krasting, S.~L. Malyshev, P.~C.~D. Milly, P.~J. Phillipps, L.~T. Sentman, B.~L. Samuels, M.~J. Spelman, M.~Winton, A.~T. Wittenberg, and N.~Zadeh.
\newblock {GFDL}’s {ESM2} global coupled climate--carbon earth system models. part {I}: Physical formulation and baseline simulation characteristics.
\newblock \emph{Journal of climate}, 25\penalty0 (19):\penalty0 6646--6665, 2012.
\newblock URL \url{https://doi.org/10.1175/JCLI-D-11-00560.1}.

\bibitem[Eyring et~al.(2019)Eyring, Cox, Flato, Gleckler, Abramowitz, Caldwell, Collins, Gier, Hall, Hoffman, Hurtt, Jahn, Jones, Klein, Krasting, Kwiatkowski, Lorenz, Maloney, Meehl, Pendergrass, Pincus, Ruane, Russell, Sanderson, Santer, Sherwood, Simpson, Stouffer, and Williamson]{eyring2019taking}
V.~Eyring, P.~M. Cox, G.~M. Flato, P.~J. Gleckler, G.~Abramowitz, P.~Caldwell, W.~D. Collins, B.~K. Gier, A.~D. Hall, F.~M. Hoffman, G.~C. Hurtt, A.~Jahn, C.~D. Jones, S.~A. Klein, J.~P. Krasting, L.~Kwiatkowski, R.~Lorenz, E.~Maloney, G.~A. Meehl, A.~G. Pendergrass, R.~Pincus, A.~C. Ruane, J.~L. Russell, B.~M. Sanderson, B.~D. Santer, S.~C. Sherwood, I.~R. Simpson, R.~J. Stouffer, and M.~S. Williamson.
\newblock Taking climate model evaluation to the next level.
\newblock \emph{Nature Climate Change}, 9\penalty0 (2):\penalty0 102--110, 2019.
\newblock URL \url{https://doi.org/10.1038/s41558-018-0355-y}.

\bibitem[Flato et~al.(2014)Flato, Marotzke, Abiodun, Braconnot, Chou, Collins, Cox, Driouech, Emori, Eyring, Forest, Gleckler, Guilyardi, Jakob, Kattsov, Reason, and Rummukainen]{flato2014evaluation}
G.~Flato, J.~Marotzke, B.~Abiodun, P.~Braconnot, S.~C. Chou, W.~Collins, P.~Cox, F.~Driouech, S.~Emori, V.~Eyring, C.~Forest, P.~Gleckler, E.~Guilyardi, C.~Jakob, V.~Kattsov, C.~Reason, and M.~Rummukainen.
\newblock Evaluation of climate models.
\newblock In \emph{Climate change 2013: the physical science basis. Contribution of Working Group I to the Fifth Assessment Report of the Intergovernmental Panel on Climate Change}, pages 741--866. Cambridge University Press, 2014.
\newblock URL \url{https://doi.org/10.1017/CBO9781107415324.020}.

\bibitem[Giorgi and Gutowski~Jr(2015)]{giorgi2015regional}
F.~Giorgi and W.~J. Gutowski~Jr.
\newblock Regional dynamical downscaling and the {CORDEX} initiative.
\newblock \emph{Annual review of environment and resources}, 40:\penalty0 467--490, 2015.
\newblock URL \url{https://doi.org/10.1146/annurev-environ-102014-021217}.

\bibitem[Giorgi et~al.(2012)Giorgi, Coppola, Solmon, Mariotti, Sylla, Bi, Elguindi, Diro, Nair, Giuliani, Turuncoglu, Cozzini, Güttler, O’Brien, Tawfik, Shalaby, Zakey, Steiner, Stordal, Sloan, and Brankovic]{giorgi2012regcm4}
F.~Giorgi, E.~Coppola, F.~Solmon, L.~Mariotti, M.~Sylla, X.~Bi, N.~Elguindi, G.~Diro, V.~Nair, G.~Giuliani, U.~U. Turuncoglu, S.~Cozzini, I.~Güttler, T.~A. O’Brien, A.~B. Tawfik, A.~Shalaby, A.~S. Zakey, A.~L. Steiner, F.~Stordal, L.~C. Sloan, and C.~Brankovic.
\newblock {RegCM4}: model description and preliminary tests over multiple {CORDEX} domains.
\newblock \emph{Climate Research}, 52:\penalty0 7--29, 2012.
\newblock URL \url{https://doi.org/10.3354/cr01018}.

\bibitem[Girona-Mata et~al.(2024)Girona-Mata, Orr, Widmann, Bannister, Dars, Hosking, Norris, Ocio, Phillips, Steiner, and Richard~E]{girona2024probabilistic}
M.~Girona-Mata, A.~Orr, M.~Widmann, D.~Bannister, G.~H. Dars, S.~Hosking, J.~Norris, D.~Ocio, T.~Phillips, J.~Steiner, and T.~Richard~E.
\newblock Probabilistic precipitation downscaling for ungauged mountain sites: a pilot study for the {Hindu Kush Karakoram Himalaya}.
\newblock \emph{EGUsphere}, 2024:\penalty0 1--33, 2024.
\newblock URL \url{https://doi.org/10.5194/egusphere-2024-2805}.

\bibitem[GTN-G(2023)]{gtn2017gtn}
GTN-G.
\newblock {GTN-G} glacier regions [{Data} set].
\newblock Global Terrestrial Network for Glaciers, 2023.
\newblock URL \url{https://doi.org/10.5904/gtng-glacreg-2023-07}.

\bibitem[Harris et~al.(2013)Harris, Sexton, Booth, Collins, and Murphy]{harris2013probabilistic}
G.~R. Harris, D.~M. Sexton, B.~B. Booth, M.~Collins, and J.~M. Murphy.
\newblock Probabilistic projections of transient climate change.
\newblock \emph{Climate dynamics}, 40:\penalty0 2937--2972, 2013.
\newblock URL \url{https://doi.org/10.1007/s00382-012-1647-y}.

\bibitem[Hausfather et~al.(2022)Hausfather, Marvel, Schmidt, Nielsen-Gammon, and Zelinka]{hausfather2022climate}
Z.~Hausfather, K.~Marvel, G.~A. Schmidt, J.~W. Nielsen-Gammon, and M.~Zelinka.
\newblock Climate simulations: Recognize the ‘hot model’ problem.
\newblock \emph{Nature}, 605\penalty0 (7908):\penalty0 26--29, 2022.
\newblock URL \url{https://doi.org/10.1038/d41586-022-01192-2}.

\bibitem[Hazeleger et~al.(2010)Hazeleger, Severijns, Semmler, Ştefănescu, Yang, Wang, Wyser, Dutra, Baldasano, Bintanja, Bougeault, Caballero, Ekman, Christensen, van~den Hurk, Jimenez, Jones, Kållberg, Koenigk, McGrath, Miranda, van Noije, Palmer, Parodi, Schmith, Selten, Storelvmo, Sterl, Tapamo, Vancoppenolle, Viterbo, and Willén]{hazeleger2010ec}
W.~Hazeleger, C.~Severijns, T.~Semmler, S.~Ştefănescu, S.~Yang, X.~Wang, K.~Wyser, E.~Dutra, J.~M. Baldasano, R.~Bintanja, P.~Bougeault, R.~Caballero, A.~M.~L. Ekman, J.~H. Christensen, B.~van~den Hurk, P.~Jimenez, C.~Jones, P.~Kållberg, T.~Koenigk, R.~McGrath, P.~Miranda, T.~van Noije, T.~Palmer, J.~A. Parodi, T.~Schmith, F.~Selten, T.~Storelvmo, A.~Sterl, H.~Tapamo, M.~Vancoppenolle, P.~Viterbo, and U.~Willén.
\newblock {EC-Earth}: a seamless earth-system prediction approach in action.
\newblock \emph{Bulletin of the American Meteorological Society}, 91\penalty0 (10):\penalty0 1357--1364, 2010.
\newblock URL \url{https://doi.org/10.1175/2010BAMS2877.1}.

\bibitem[Hersbach(2000)]{hersbach2000decomposition}
H.~Hersbach.
\newblock Decomposition of the continuous ranked probability score for ensemble prediction systems.
\newblock \emph{Weather and Forecasting}, 15\penalty0 (5):\penalty0 559 -- 570, 2000.
\newblock \doi{10.1175/1520-0434(2000)015<0559:DOTCRP>2.0.CO;2}.
\newblock URL \url{https://journals.ametsoc.org/view/journals/wefo/15/5/1520-0434_2000_015_0559_dotcrp_2_0_co_2.xml}.

\bibitem[{ICIMOD}(2008)]{icimod2008outline}
{ICIMOD}.
\newblock Outline boundary of {Hindu Kush Himalayan} ({HKH}) region [data set].
\newblock International Centre for Integrated Mountain Development (ICIMOD), 2008.
\newblock URL \url{https://doi.org/10.26066/RDS.3924}.

\bibitem[Immerzeel et~al.(2020)Immerzeel, Lutz, Andrade, Bahl, Biemans, Bolch, Hyde, Brumby, Davies, Elmore, Emmer, Feng, Fern{\'a}ndez, Haritashya, Kargel, Koppes, Kraaijenbrink, Kulkarni, Mayewski, Nepal, Pacheco, Painter, Pellicciotti, Rajaram, Rupper, Sinisalo, Shrestha, Viviroli, Wada, Xiao, dong Yao, and Baillie]{immerzeel2020importance}
W.~W. Immerzeel, A.~F. Lutz, M.~Andrade, A.~Bahl, H.~Biemans, T.~Bolch, S.~Hyde, S.~P. Brumby, B.~J. Davies, A.~C. Elmore, A.~Emmer, M.~Feng, A.~Fern{\'a}ndez, U.~K. Haritashya, J.~S. Kargel, M.~N. Koppes, P.~D.~A. Kraaijenbrink, A.~V. Kulkarni, P.~A. Mayewski, S.~Nepal, P.~Pacheco, T.~H. Painter, F.~Pellicciotti, H.~Rajaram, S.~B. Rupper, A.~Sinisalo, A.~B. Shrestha, D.~Viviroli, Y.~Wada, C.~Xiao, T.~dong Yao, and J.~E.~M. Baillie.
\newblock Importance and vulnerability of the world’s water towers.
\newblock \emph{Nature}, 577\penalty0 (7790):\penalty0 364--369, 2020.
\newblock URL \url{https://doi.org/10.1038/s41586-019-1822-y}.

\bibitem[Jeffrey et~al.(2013)Jeffrey, Rotstayn, Collier, Dravitzki, Hamalainen, Moeseneder, Wong, and Syktus]{jeffrey2013australia}
S.~Jeffrey, L.~Rotstayn, M.~Collier, S.~Dravitzki, C.~Hamalainen, C.~Moeseneder, K.~Wong, and J.~Syktus.
\newblock {Australia}’s {CMIP5} submission using the {CSIRO-Mk3.6} model.
\newblock \emph{Australian Meteorological and Oceanographic Journal}, 63\penalty0 (1):\penalty0 1--13, 2013.
\newblock URL \url{https://doi.org/10.1071/ES13001}.

\bibitem[Jiang et~al.(2023)Jiang, Zhou, Qian, Li, Song, Li, Chen, Zhang, and Chen]{jiang2023precipitation}
J.~Jiang, T.~Zhou, Y.~Qian, C.~Li, F.~Song, H.~Li, X.~Chen, W.~Zhang, and Z.~Chen.
\newblock Precipitation regime changes in {High Mountain Asia} driven by cleaner air.
\newblock \emph{Nature}, 623\penalty0 (7987):\penalty0 544--549, 2023.
\newblock URL \url{https://doi.org/10.1038/s41586-023-06619-y}.

\bibitem[Jungclaus et~al.(2013)Jungclaus, Fischer, Haak, Lohmann, Marotzke, Matei, Mikolajewicz, Notz, and Von~Storch]{jungclaus2013characteristics}
J.~H. Jungclaus, N.~Fischer, H.~Haak, K.~Lohmann, J.~Marotzke, D.~Matei, U.~Mikolajewicz, D.~Notz, and J.~Von~Storch.
\newblock Characteristics of the ocean simulations in the {Max Planck Institute Ocean Model (MPIOM)} the ocean component of the {MPI-Earth} system model.
\newblock \emph{Journal of Advances in Modeling Earth Systems}, 5\penalty0 (2):\penalty0 422--446, 2013.
\newblock URL \url{https://doi.org/10.1002/jame.20023}.

\bibitem[Kharin and Zwiers(2002)]{kharin2002climate}
V.~V. Kharin and F.~W. Zwiers.
\newblock Climate predictions with multimodel ensembles.
\newblock \emph{Journal of Climate}, 15\penalty0 (7):\penalty0 793--799, 2002.
\newblock URL \url{https://doi.org/10.1175/1520-0442(2002)015<0793:CPWME>2.0.CO;2}.

\bibitem[Knutti(2010)]{knutti2010end}
R.~Knutti.
\newblock The end of model democracy? an editorial comment.
\newblock \emph{Climatic change}, 102\penalty0 (3):\penalty0 395--404, 2010.
\newblock URL \url{https://doi.org/10.1007/s10584-010-9800-2}.

\bibitem[Lee et~al.(2021)Lee, Marotzke, Bala, Cao, Corti, Dunne, Engelbrecht, Fischer, Fyfe, Jones, Maycock, Joseph, Ndiaye, Swapna, and Zhou]{lee2021future}
J.-Y. Lee, J.~Marotzke, G.~Bala, L.~Cao, S.~Corti, J.~P. Dunne, F.~Engelbrecht, E.~Fischer, J.~C. Fyfe, C.~Jones, A.~Maycock, M.~Joseph, O.~Ndiaye, P.~Swapna, and T.~Zhou.
\newblock Future global climate: scenario-based projections and near-term information.
\newblock In \emph{Climate change 2021: The physical science basis. Contribution of working group I to the Sixth Assessment Report of the Intergovernmental Panel on Climate Change}, pages 553--672. Cambridge University Press, 2021.
\newblock URL \url{https://doi.org/10.1017/9781009157896.006}.

\bibitem[Lehner and Grill(2013)]{lehner2013global}
B.~Lehner and G.~Grill.
\newblock Global river hydrography and network routing: baseline data and new approaches to study the world's large river systems.
\newblock \emph{Hydrological Processes}, 27\penalty0 (15):\penalty0 2171--2186, 2013.
\newblock URL \url{http://doi.org/10.1002/hyp.9740}.

\bibitem[Martinez-Villalobos and Neelin(2021)]{martinez2021climate}
C.~Martinez-Villalobos and J.~D. Neelin.
\newblock Climate models capture key features of extreme precipitation probabilities across regions.
\newblock \emph{Environmental Research Letters}, 16\penalty0 (2):\penalty0 024017, 2021.
\newblock URL \url{https://doi.org/10.1088/1748-9326/abd351}.

\bibitem[Masson and Knutti(2011)]{masson2011climate}
D.~Masson and R.~Knutti.
\newblock Climate model genealogy.
\newblock \emph{Geophysical Research Letters}, 38\penalty0 (8), 2011.
\newblock URL \url{https://doi.org/10.1029/2011GL046864}.

\bibitem[Massoud et~al.(2023)Massoud, Lee, Terando, and Wehner]{massoud2023bayesian}
E.~C. Massoud, H.~K. Lee, A.~Terando, and M.~Wehner.
\newblock {Bayesian} weighting of climate models based on climate sensitivity.
\newblock \emph{Communications Earth \& Environment}, 4\penalty0 (1):\penalty0 365, 2023.
\newblock URL \url{https://doi.org/10.1038/s43247-023-01009-8}.

\bibitem[Maussion et~al.(2014)Maussion, Scherer, M{\"o}lg, Collier, Curio, and Finkelnburg]{maussion2014precipitation}
F.~Maussion, D.~Scherer, T.~M{\"o}lg, E.~Collier, J.~Curio, and R.~Finkelnburg.
\newblock Precipitation seasonality and variability over the {Tibetan Plateau} as resolved by the {High Asia Reanalysis}.
\newblock \emph{Journal of Climate}, 27\penalty0 (5):\penalty0 1910--1927, 2014.
\newblock URL \url{https://doi.org/10.1175/JCLI-D-13-00282.1}.

\bibitem[Min et~al.(2007)Min, Simonis, and Hense]{min2007probabilistic}
S.-K. Min, D.~Simonis, and A.~Hense.
\newblock Probabilistic climate change predictions applying {Bayesian} model averaging.
\newblock \emph{Philosophical Transactions of the Royal Society A: Mathematical, Physical and Engineering Sciences}, 365\penalty0 (1857):\penalty0 2103--2116, 2007.
\newblock \doi{https://doi.org/10.1098/rsta.2007.2070}.

\bibitem[Nie et~al.(2021)Nie, Pritchard, Liu, Hennig, Wang, Wang, Liu, Nepal, Samyn, Hewitt, et~al.]{nie2021glacial}
Y.~Nie, H.~D. Pritchard, Q.~Liu, T.~Hennig, W.~Wang, X.~Wang, S.~Liu, S.~Nepal, D.~Samyn, K.~Hewitt, et~al.
\newblock Glacial change and hydrological implications in the {Himalaya} and {Karakoram}.
\newblock \emph{Nature reviews earth \& environment}, 2\penalty0 (2):\penalty0 91--106, 2021.
\newblock URL \url{https://doi.org/10.1038/s43017-020-00124-w}.

\bibitem[Norris et~al.(2017)Norris, Carvalho, Jones, Cannon, Bookhagen, Palazzi, and Tahir]{norris2017spatiotemporal}
J.~Norris, L.~M. Carvalho, C.~Jones, F.~Cannon, B.~Bookhagen, E.~Palazzi, and A.~A. Tahir.
\newblock The spatiotemporal variability of precipitation over the {Himalaya}: evaluation of one-year {WRF} model simulation.
\newblock \emph{Climate Dynamics}, 49:\penalty0 2179--2204, 2017.
\newblock URL \url{https://doi.org/10.1007/s00382-016-3414-y}.

\bibitem[Norris et~al.(2019)Norris, Carvalho, Jones, and Cannon]{norris2019deciphering}
J.~Norris, L.~M. Carvalho, C.~Jones, and F.~Cannon.
\newblock Deciphering the contrasting climatic trends between the central {Himalaya }and {Karakoram} with 36 years of {WRF} simulations.
\newblock \emph{Climate Dynamics}, 52:\penalty0 159--180, 2019.
\newblock URL \url{https://doi.org/10.1007/s00382-018-4133-3}.

\bibitem[Norris et~al.(2020)Norris, Carvalho, Jones, and Cannon]{norris2020warming}
J.~Norris, L.~M. Carvalho, C.~Jones, and F.~Cannon.
\newblock Warming and drying over the central {Himalaya} caused by an amplification of local mountain circulation.
\newblock \emph{NPJ Climate and Atmospheric Science}, 3\penalty0 (1):\penalty0 1, 2020.
\newblock URL \url{https://doi.org/10.1038/s41612-019-0105-5}.

\bibitem[Olson et~al.(2016)Olson, Fan, and Evans]{olson2016simple}
R.~Olson, Y.~Fan, and J.~P. Evans.
\newblock A simple method for {Bayesian} model averaging of regional climate model projections: Application to southeast {Australian} temperatures.
\newblock \emph{Geophysical Research Letters}, 43\penalty0 (14):\penalty0 7661--7669, 2016.
\newblock \doi{https://doi.org/10.1002/2016GL069704}.

\bibitem[Orr et~al.(2017)Orr, Listowski, Couttet, Collier, Immerzeel, Deb, and Bannister]{orr2017sensitivity}
A.~Orr, C.~Listowski, M.~Couttet, E.~Collier, W.~Immerzeel, P.~Deb, and D.~Bannister.
\newblock Sensitivity of simulated summer monsoonal precipitation in {Langtang Valley}, {Himalaya}, to cloud microphysics schemes in {WRF}.
\newblock \emph{Journal of Geophysical Research: Atmospheres}, 122\penalty0 (12):\penalty0 6298--6318, 2017.
\newblock URL \url{https://doi.org/10.1002/2016JD025801}.

\bibitem[Orr et~al.(2022)Orr, Ahmad, Alam, Appadurai, Bharucha, Biemans, Bolch, Chaulagain, Dhaubanjar, Dimri, Dixon, Fowler, Gioli, Halvorson, Hussain, Jeelani, Kamal, Khalid, Liu, Lutz, Mehra, Miles, Momblanch, Muccione, Mukherji, Mustafa, Najmuddin, Nasimi, Nüsser, Pandey, Parveen, Pellicciotti, Pollino, Potter, Qazizada, Ray, Romshoo, Sarkar, Sawas, Sen, Shah, Shah, Shea, Sheikh, Shrestha, Tayal, Tigala, Virk, Wester, and Wescoat~Jr.]{orr2022knowledge}
A.~Orr, B.~Ahmad, U.~Alam, A.~Appadurai, Z.~P. Bharucha, H.~Biemans, T.~Bolch, N.~P. Chaulagain, S.~Dhaubanjar, A.~P. Dimri, H.~Dixon, H.~J. Fowler, G.~Gioli, S.~J. Halvorson, A.~Hussain, G.~Jeelani, S.~Kamal, I.~S. Khalid, S.~Liu, A.~Lutz, M.~K. Mehra, E.~Miles, A.~Momblanch, V.~Muccione, A.~Mukherji, D.~Mustafa, O.~Najmuddin, M.~N. Nasimi, M.~Nüsser, V.~P. Pandey, S.~Parveen, F.~Pellicciotti, C.~Pollino, E.~Potter, M.~R. Qazizada, S.~Ray, S.~Romshoo, S.~K. Sarkar, A.~Sawas, S.~Sen, A.~Shah, M.~A.~A. Shah, J.~M. Shea, A.~T. Sheikh, A.~B. Shrestha, S.~Tayal, S.~Tigala, Z.~T. Virk, P.~Wester, and J.~L. Wescoat~Jr.
\newblock Knowledge priorities on climate change and water in the {Upper Indus Basin}: A horizon scanning exercise to identify the top 100 research questions in social and natural sciences.
\newblock \emph{Earth's Future}, 10\penalty0 (4):\penalty0 e2021EF002619, 2022.
\newblock URL \url{https://doi.org/10.1029/2021EF002619}.

\bibitem[Palazzi et~al.(2015)Palazzi, von Hardenberg, Terzago, and Provenzale]{palazzi2015precipitation}
E.~Palazzi, J.~von Hardenberg, S.~Terzago, and A.~Provenzale.
\newblock Precipitation in the {Karakoram-Himalaya}: a {CMIP5} view.
\newblock \emph{Climate Dynamics}, 45:\penalty0 21--45, 2015.
\newblock URL \url{https://doi.org/10.1007/s00382-014-2341-z}.

\bibitem[Palmer et~al.(2005)Palmer, Doblas-Reyes, Hagedorn, and Weisheimer]{palmer2005probabilistic}
T.~Palmer, F.~Doblas-Reyes, R.~Hagedorn, and A.~Weisheimer.
\newblock Probabilistic prediction of climate using multi-model ensembles: from basics to applications.
\newblock \emph{Philosophical Transactions of the Royal Society B: Biological Sciences}, 360\penalty0 (1463):\penalty0 1991--1998, 2005.
\newblock URL \url{https://doi.org/10.1098/rstb.2005.1750}.

\bibitem[Panaretos and Zemel(2019)]{panaretos2019statistical}
V.~M. Panaretos and Y.~Zemel.
\newblock Statistical aspects of {Wasserstein} distances.
\newblock \emph{Annual review of statistics and its application}, 6\penalty0 (1):\penalty0 405--431, 2019.
\newblock URL \url{https://doi.org/10.1146/annurev-statistics-030718-104938}.

\bibitem[Panday et~al.(2015)Panday, Thibeault, and Frey]{panday2015cmip}
P.~K. Panday, J.~Thibeault, and K.~E. Frey.
\newblock Changing temperature and precipitation extremes in the {Hindu Kush-Himalayan} region: an analysis of {CMIP3} and {CMIP5} simulations and projections.
\newblock \emph{International Journal of Climatology}, 35\penalty0 (10):\penalty0 3058--3077, 2015.
\newblock \doi{https://doi.org/10.1002/joc.4192}.
\newblock URL \url{https://rmets.onlinelibrary.wiley.com/doi/abs/10.1002/joc.4192}.

\bibitem[Rasmussen and Williams(2006)]{rasmussen2006gaussian}
C.~E. Rasmussen and C.~K. Williams.
\newblock \emph{{Gaussian processes} for machine learning}, volume~1.
\newblock MIT Press, 2006.
\newblock URL \url{https://doi.org/10.7551/mitpress/3206.001.0001}.

\bibitem[{RGI Consortium}(2017)]{consortium2017randolph}
{RGI Consortium}.
\newblock {Randolph Glacier Inventory}--a dataset of global glacier outlines: Version 6.0 [data set].
\newblock National Snow and Ice Data Center, 2017.
\newblock URL \url{https://doi.org/10.7265/4m1f-gd79}.

\bibitem[Samuelsson et~al.(2015)Samuelsson, Gollvik, Kupiainen, Kourzeneva, and van~de Berg]{samuelsson2015surface}
P.~Samuelsson, S.~Gollvik, M.~Kupiainen, E.~Kourzeneva, and W.~J. van~de Berg.
\newblock The surface processes of the {Rossby Centre} regional atmospheric climate model ({RCA4}).
\newblock Technical report, SMHI, 2015.
\newblock URL \url{https://www.smhi.se/polopoly_fs/1.89799!/Menu/general/extGroup/attachmentColHold/mainCol1/file/meteorologi_157.pdf}.
\newblock Accessed 18 November 2024.

\bibitem[Sanderson and Knutti(2012)]{sanderson2012interpretation}
B.~M. Sanderson and R.~Knutti.
\newblock On the interpretation of constrained climate model ensembles.
\newblock \emph{Geophysical Research Letters}, 39\penalty0 (16), 2012.
\newblock URL \url{https://doi.org/10.1029/2012GL052665}.

\bibitem[Sanjay et~al.(2017)Sanjay, Krishnan, Shrestha, Rajbhandari, and Ren]{sanjay2017downscaled}
J.~Sanjay, R.~Krishnan, A.~B. Shrestha, R.~Rajbhandari, and G.-Y. Ren.
\newblock Downscaled climate change projections for the {Hindu Kush Himalayan} region using {CORDEX South Asia} regional climate models.
\newblock \emph{Advances in Climate Change Research}, 8\penalty0 (3):\penalty0 185--198, 2017.
\newblock URL \url{https://doi.org/10.1016/j.accre.2017.08.003}.

\bibitem[{Sanjay Jayanarayanan, Scientist}(2025)]{grid-email}
{Sanjay Jayanarayanan, Scientist}.
\newblock Centre for climate change research at indian institute of tropical meteorology.
\newblock Personal correspondence, May 2025.

\bibitem[Saul et~al.(2016)Saul, Hensman, Vehtari, and Lawrence]{saul2016chained}
A.~D. Saul, J.~Hensman, A.~Vehtari, and N.~D. Lawrence.
\newblock Chained {Gaussian} processes.
\newblock In \emph{Proceedings of the 19th International Conference on Artificial Intelligence and Statistics}, pages 1431--1440, 09--11 May 2016.

\bibitem[Snelson et~al.(2003)Snelson, Ghahramani, and Rasmussen]{snelson2003warped}
E.~Snelson, Z.~Ghahramani, and C.~Rasmussen.
\newblock Warped {Gaussian} processes.
\newblock In \emph{Advances in Neural Information Processing Systems}, volume~16, Vacouver, BC, Canada, 8--13 December 2003.
\newblock URL \url{https://papers.nips.cc/paper_files/paper/2003/file/6b5754d737784b51ec5075c0dc437bf0-Paper.pdf}.
\newblock Accessed 11 November 2024.

\bibitem[Stainforth et~al.(2007)Stainforth, Downing, Washington, Lopez, and New]{stainforth2007issues}
D.~A. Stainforth, T.~E. Downing, R.~Washington, A.~Lopez, and M.~New.
\newblock Issues in the interpretation of climate model ensembles to inform decisions.
\newblock \emph{Philosophical Transactions of the Royal Society A: Mathematical, Physical and Engineering Sciences}, 365\penalty0 (1857):\penalty0 2163--2177, 2007.
\newblock URL \url{https://doi.org/10.1098/rsta.2007.2073}.

\bibitem[Taylor et~al.(2012)Taylor, Stouffer, and Meehl]{taylor2012overview}
K.~E. Taylor, R.~J. Stouffer, and G.~A. Meehl.
\newblock An overview of {CMIP5} and the experiment design.
\newblock \emph{Bulletin of the American meteorological Society}, 93\penalty0 (4):\penalty0 485--498, 2012.
\newblock URL \url{https://doi.org/10.1175/BAMS-D-11-00094.1}.

\bibitem[Tazi et~al.(2024)Tazi, Orr, Hernandez-González, Hosking, and Turner]{tazi2024downscaling}
K.~Tazi, A.~O. Orr, J.~Hernandez-González, S.~Hosking, and R.~E. Turner.
\newblock Downscaling precipitation over {High-mountain Asia} using multi-fidelity {Gaussian} processes: improved estimates from {ERA5}.
\newblock \emph{Hydrology and Earth System Sciences}, 28\penalty0 (22):\penalty0 4903--4925, 2024.
\newblock URL \url{https://doi.org/10.5194/hess-28-4903-2024}.

\bibitem[Tazi et~al.(2025)Tazi, Orr, Hosking, and Turner]{tazi2024precipitation}
K.~Tazi, A.~O. Orr, S.~Hosking, and R.~E. Turner.
\newblock Precipitation prediction from large-scale climatic features over the {Upper Indus Basin} using non-stationary {Gaussian} processes.
\newblock \emph{Environmental Data Science}, 2025.
\newblock Accepted.

\bibitem[Tebaldi and Knutti(2007)]{tebaldi2007use}
C.~Tebaldi and R.~Knutti.
\newblock The use of the multi-model ensemble in probabilistic climate projections.
\newblock \emph{Philosophical transactions of the royal society A: mathematical, physical and engineering sciences}, 365\penalty0 (1857):\penalty0 2053--2075, 2007.
\newblock URL \url{https://doi.org/10.1098/rsta.2007.2076}.

\bibitem[Teichmann et~al.(2013)Teichmann, Eggert, Elizalde, Haensler, Jacob, Kumar, Moseley, Pfeifer, Rechid, Remedio, Ries, Petersen, Preuschmann, Raub, Saeed, Sieck, and Weber]{teichmann2013does}
C.~Teichmann, B.~Eggert, A.~Elizalde, A.~Haensler, D.~Jacob, P.~Kumar, C.~Moseley, S.~Pfeifer, D.~Rechid, A.~R. Remedio, H.~Ries, J.~Petersen, S.~Preuschmann, T.~Raub, F.~Saeed, K.~Sieck, and T.~Weber.
\newblock How does a regional climate model modify the projected climate change signal of the driving {GCM}: a study over different {CORDEX} regions using remo.
\newblock \emph{Atmosphere}, 4\penalty0 (2):\penalty0 214--236, 2013.
\newblock URL \url{https://doi.org/10.3390/atmos4020214}.

\bibitem[T{\'e}l et~al.(2020)T{\'e}l, B{\'o}dai, Dr{\'o}tos, Haszpra, Herein, Kasz{\'a}s, and Vincze]{tel2020theory}
T.~T{\'e}l, T.~B{\'o}dai, G.~Dr{\'o}tos, T.~Haszpra, M.~Herein, B.~Kasz{\'a}s, and M.~Vincze.
\newblock The theory of parallel climate realizations: A new framework of ensemble methods in a changing climate: An overview.
\newblock \emph{Journal of Statistical Physics}, 179:\penalty0 1496--1530, 2020.
\newblock URL \url{https://doi.org/10.1007/s10955-019-02445-7}.

\bibitem[Titsias(2009)]{titsias2009variational}
M.~Titsias.
\newblock Variational learning of inducing variables in sparse {Gaussian} processes.
\newblock In \emph{Proceedings of the Twelfth International Conference on Artificial Intelligence and Statistics}, volume~5, pages 567--574, Clearwater Beach, FL, USA, 16--18 Apr 2009. PMLR.
\newblock URL \url{http://proceedings.mlr.press/v5/titsias09a/titsias09a.pdf}.
\newblock Accessed 11 November 2024.

\bibitem[Tresp(2000)]{tresp2000bayesian}
V.~Tresp.
\newblock A {Bayesian} committee machine.
\newblock \emph{Neural computation}, 12\penalty0 (11):\penalty0 2719--2741, 2000.
\newblock URL \url{https://doi.org/10.1162/089976600300014908}.

\bibitem[Voldoire et~al.(2013)Voldoire, Sanchez-Gomez, Salas~y M{\'e}lia, Decharme, Cassou, S{\'e}n{\'e}si, Valcke, Beau, Alias, Chevallier, Déqué, Deshayes, Douville, Fernandez, Madec, Maisonnave, Moine, Planton, Saint-Martin, Szopa, Tyteca, Alkama, Belamari, Braun, Coquart, and Chauvin]{voldoire2013cnrm}
A.~Voldoire, E.~Sanchez-Gomez, D.~Salas~y M{\'e}lia, B.~Decharme, C.~Cassou, S.~S{\'e}n{\'e}si, S.~Valcke, I.~Beau, A.~Alias, M.~Chevallier, M.~Déqué, J.~Deshayes, H.~Douville, E.~Fernandez, G.~Madec, E.~Maisonnave, M.-P. Moine, S.~Planton, D.~Saint-Martin, S.~Szopa, S.~Tyteca, R.~Alkama, S.~Belamari, A.~Braun, L.~Coquart, and F.~Chauvin.
\newblock The {CNRM-CM5.1} global climate model: description and basic evaluation.
\newblock \emph{Climate dynamics}, 40:\penalty0 2091--2121, 2013.
\newblock URL \url{https://doi.org/10.1007/s00382-011-1259-y}.

\bibitem[Watanabe et~al.(2010)Watanabe, Suzuki, O’ishi, Komuro, Watanabe, Emori, Takemura, Chikira, Ogura, Sekiguchi, Takata, Yamazaki, Yokohata, Nozawa, Hasumi, Tatebe, and Kimoto]{watanabe2010improved}
M.~Watanabe, T.~Suzuki, R.~O’ishi, Y.~Komuro, S.~Watanabe, S.~Emori, T.~Takemura, M.~Chikira, T.~Ogura, M.~Sekiguchi, K.~Takata, D.~Yamazaki, T.~Yokohata, T.~Nozawa, H.~Hasumi, H.~Tatebe, and M.~Kimoto.
\newblock Improved climate simulation by {MIROC5}: mean states, variability, and climate sensitivity.
\newblock \emph{Journal of Climate}, 23\penalty0 (23):\penalty0 6312--6335, 2010.
\newblock URL \url{https://doi.org/10.1175/2010JCLI3679.1}.

\bibitem[Wester et~al.(2023)Wester, Chaudhary, Chettri, Maharjan, Nepal, and Steiner]{wester2023water}
P.~Wester, S.~Chaudhary, N.~Chettri, A.~Maharjan, S.~Nepal, and J.~Steiner.
\newblock \emph{Water, ice, society, and ecosystems in the {Hindu Kush Himalaya}: an outlook}.
\newblock International Centre for Integrated Mountain Development, Kathmandu, Nepal, 2023.
\newblock URL \url{https//doi.org/10.53055/ICIMOD.1028}.

\bibitem[Westermann et~al.(2024)Westermann, Barboux, Bartsch, Delaloye, Grosse, Heim, Hugelius, Irrgang, Kääb, Matthes, Nitze, Pellet, Seifert, Strozzi, Wegmüller, Wieczorek, and Wiesmann]{westermann2024esa}
S.~Westermann, C.~Barboux, A.~Bartsch, R.~Delaloye, G.~Grosse, B.~Heim, G.~Hugelius, A.~Irrgang, A.~Kääb, H.~Matthes, I.~Nitze, C.~Pellet, F.~Seifert, T.~Strozzi, U.~Wegmüller, M.~Wieczorek, and A.~Wiesmann.
\newblock {ESA Permafrost Climate Change Initiative}: Permafrost version 4 data products [data set].
\newblock \emph{NERC EDS Centre for Environmental Data Analysis}, 2024.
\newblock URL \url{https://doi.org/10.5285/93444bc1c4364a59869e004bf9bfd94a}.

\bibitem[Wilcox et~al.(2020)Wilcox, Liu, Samset, Hawkins, Lund, Nordling, Undorf, Bollasina, Ekman, Krishnan, et~al.]{wilcox2020accelerated}
L.~J. Wilcox, Z.~Liu, B.~H. Samset, E.~Hawkins, M.~T. Lund, K.~Nordling, S.~Undorf, M.~Bollasina, A.~M. Ekman, S.~Krishnan, et~al.
\newblock Accelerated increases in global and {Asian} summer monsoon precipitation from future aerosol reductions.
\newblock \emph{Atmospheric Chemistry and Physics}, 20\penalty0 (20):\penalty0 11955--11977, 2020.
\newblock URL \url{https://doi.org/10.3390/atmos9030080}.

\bibitem[Yatagai et~al.(2012)Yatagai, Kamiguchi, Arakawa, Hamada, Yasutomi, and Kitoh]{yatagai2012aphrodite}
A.~Yatagai, K.~Kamiguchi, O.~Arakawa, A.~Hamada, N.~Yasutomi, and A.~Kitoh.
\newblock {APHRODITE}: Constructing a long-term daily gridded precipitation dataset for {Asia} based on a dense network of rain gauges.
\newblock \emph{Bulletin of the American Meteorological Society}, 93\penalty0 (9):\penalty0 1401--1415, 2012.
\newblock URL \url{https://doi.org/10.1175/BAMS-D-11-00122.1}.

\bibitem[Zelinka et~al.(2020)Zelinka, Myers, McCoy, Po-Chedley, Caldwell, Ceppi, Klein, and Taylor]{zelinka2020causes}
M.~D. Zelinka, T.~A. Myers, D.~T. McCoy, S.~Po-Chedley, P.~M. Caldwell, P.~Ceppi, S.~A. Klein, and K.~E. Taylor.
\newblock Causes of higher climate sensitivity in {CMIP6} models.
\newblock \emph{Geophysical Research Letters}, 47\penalty0 (1):\penalty0 e2019GL085782, 2020.
\newblock URL \url{https://doi.org/10.1029/2019GL085782}.

\end{thebibliography}

\appendix

% \section{Wasserstein variances}
% \label{app:wass_var}

% \begin{figure}[ht]
% \centering
% \includegraphics[width=\textwidth]{figures/rcm_agreement.pdf}
% \caption{Wasserstein distance variance across RCM ensemble.}
% \end{figure}

% \section{CRPS differences by RCM}
% \label{app:crps_rcm_diff}

% \begin{figure}[ht]
% \centering
% \includegraphics[width=\textwidth]{figures/moe_rcm_crps_alpha_nop95_1981_2005.pdf}
% \caption{Annual-averaged CRPS difference between MoE and RCM models. Negative values (blue) imply MoE matches APHRODITE more closely while positive values (red) represent location where the RCM performs better. The plot shows the MoE generally improves over choosing one single model over the held-out validation period (1981--2005).}
% \end{figure}

\clearpage

\section{BCM softmax-variance control}
\label{app:softmax-variance}
Following \citet{cohen2020healing}, we use their general expression for expert weights:
\begin{equation}
    \beta_j(\bm{x}_*) \propto \exp(-1/\psi_j(\bm{x}_*)\tau), \quad
    \sum^{M}_{j=1} \beta_j(\bm{x}_*)= 1,
\end{equation}
where $\tau$ is a temperature parameter (with a similar function to the variable $T$ in \cref{sec:across_rcms}) that controls the sparsity between experts by multiplicatively compounding the weights of stronger experts. The function $\psi_j(\bm{x}_*)$
describes the level of confidence of the $j$\textsuperscript{th} expert at test point $\bm{x}_*$. Like \citet{cohen2020healing}, we
set $\psi_j(\bm{x}_*)$ to the posterior predictive variance at $\bm{x}_*$, i.e., $\sigma^2_j(\bm{x}_*)$ and choose $\tau=1/8$.

\section{MoE probability distribution} \label{apdx:moe-dist}
The transformed precipitation $\tilde y$ for the $r$\textsuperscript{th} RCM surrogate at some input coordinate $\bm{x}_*$ is assumed to be normally distributed. Suppressing input variables and fixed hyperparameters, we have
\begin{align}
    p_r(\tilde y_r) = \frac{1}{\sqrt{2 \pi \sigma_r^2}} \exp\left(- \frac{(\tilde y_r - \mu_r)^2}{2 \sigma_r^2}\right).
\end{align}
The distribution for $y$ can then be found by transforming the coordinates and calculating the Jacobian $p_r(\tilde y_r(y)) \frac{d \tilde y_r}{d y}$, yielding
\begin{align}
    p'_r(y) = \frac{y^{\lambda_r - 1}}{\sqrt{2 \pi \sigma_r^2}} \exp\left(-\frac{\left(\frac{y^{\lambda_r}-1}{\lambda_r} - \mu_r\right)^2}{2 \sigma_r^2}\right).
\end{align}
However, there is a small issue with this procedure. Inverting the transformation, we find that for $\lambda_r \neq 0$, $y(\tilde y) = (\lambda_r \tilde y + 1)^{1/\lambda_r}$. This quantity is ambiguous for $\tilde y_r < -1/\lambda_r$, as the fractional power of a negative number is ill-defined. To remedy this, we restrict the domain of $\tilde y$ to $(-1/\lambda_r, \infty)$, with the cost that we need to renormalise the probability distribution by its integral. Defining
\begin{align}
    C_r := \int_{-1/\lambda_r}^\infty d \tilde y_r p'_r(\tilde y_r) = \int_{-(\lambda_r^{-1} + \mu_r)/\sigma_r}^\infty \frac{1}{\sqrt{2 \pi}} e^{-x^2/2} = \frac{1}{2} \left(1 + \mathrm{erf}\left[\frac{1}{\sqrt{2 \sigma_r^2}}\left(\frac{1}{\lambda_r} + \mu_r \right)\right]\right),
\end{align}
the properly normalised probability distribution for $y$ is given by
\begin{align}
    p_r(y) = \frac{y^{\lambda_r - 1}}{C_r \sqrt{2 \pi \sigma_r^2}} \exp\left(-\frac{\left(\frac{y^{\lambda_r}-1}{\lambda_r} - \mu_r\right)^2}{2 \sigma_r^2}\right),
\end{align}
with domain $y \in [0, \infty)$. In practice, we expect $C_r$ to be very close to one.

Finally, given some weights $\{w_r\}_r$, the MoE's distribution is given by
\begin{align}
    p(y) = \sum_r w_r p_r(y),
\end{align}
with the constraint that $\sum_r w_r = 1$.

\clearpage

\section{Which RCM, when and where?}
\label{sec:best_rcm}

\begin{figure}[ht]
    \center
    \includegraphics[width=\textwidth]{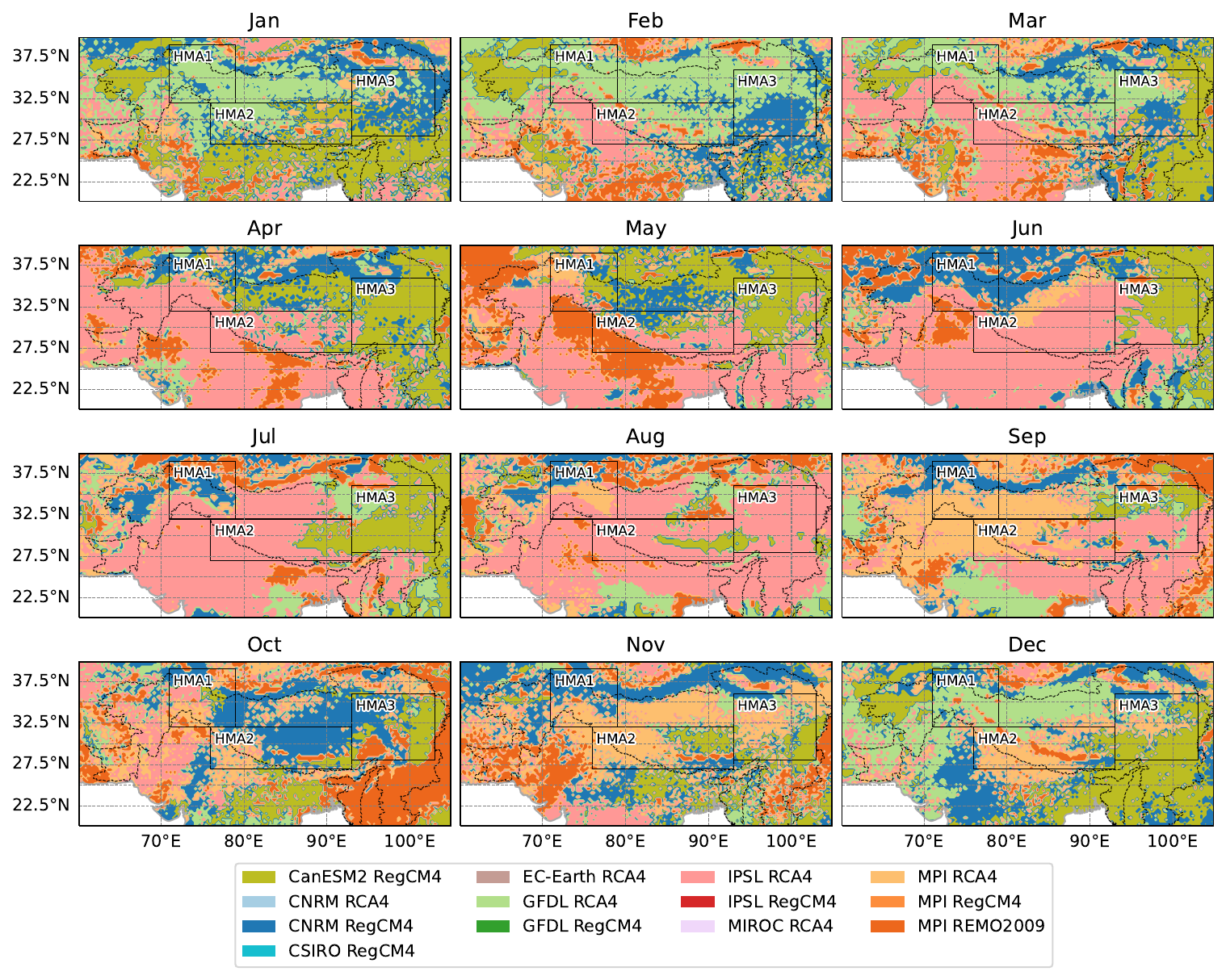}
\caption[RCM surrogates with the smallest Wasserstein distances]{RCMs with the smallest Wasserstein distances for each month and location over HMA. The distances are calculated using the scaled distributions of precipitation between 1951 and 2005. The spatial distribution of these models is not random but follows distinct spatiotemporal patterns.}
\label{fig:w_dist}
\end{figure}

\begin{figure}[ht]
\centering
\includegraphics[width=\textwidth]{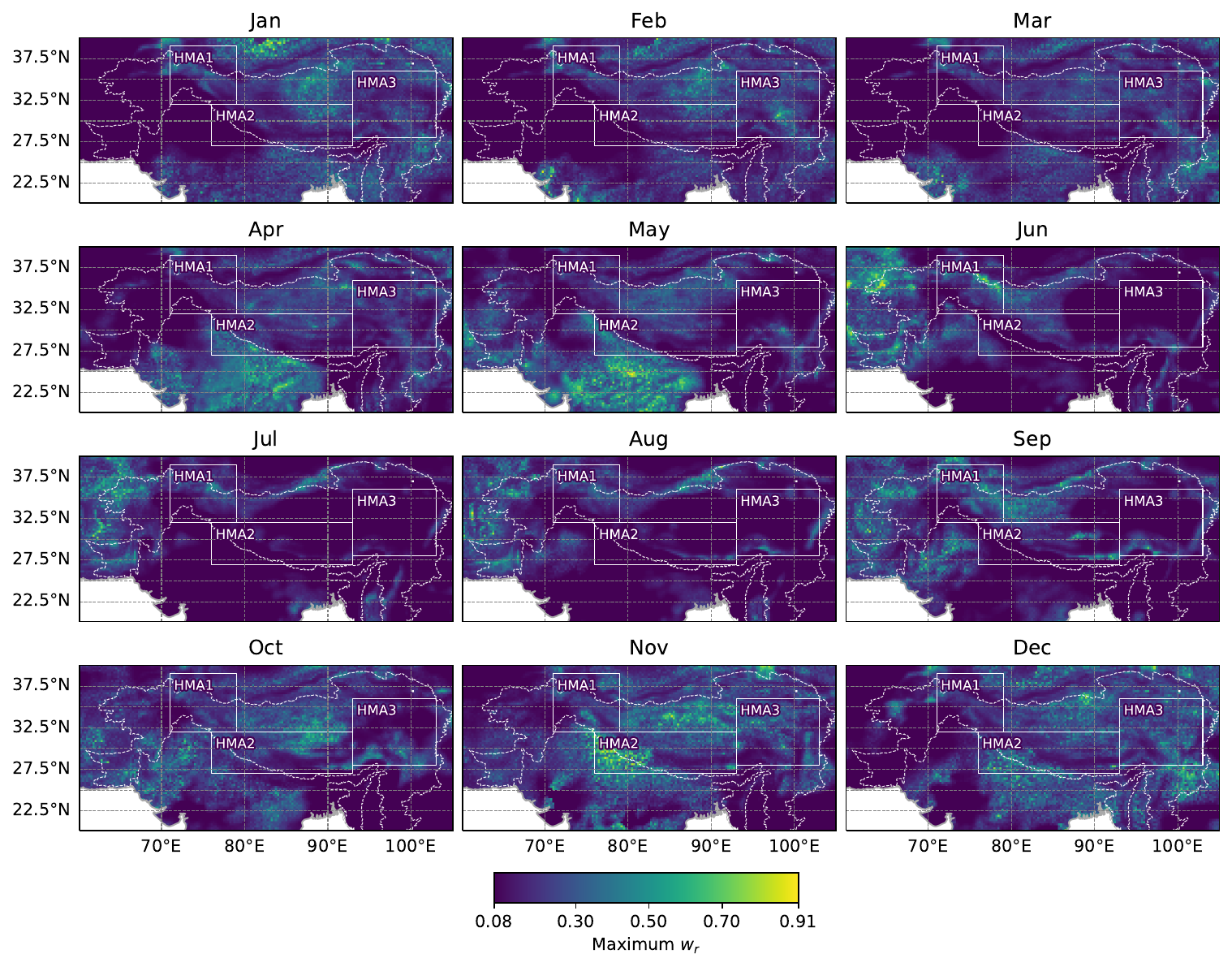}
\caption[Maximum MoE weights]{Maximum MoE weights for each month and location over HMA. The weights are optimised between 1951 and 2005. Here, one RCM surrogate (BCM) is never chosen over all others, i.e. the maximum $w_r$ is never 1. The maximum BCM weighting is highest over the Hindu Kush during the summer and over the Tibetan Plateau during the winter. Conversely, the MoE with close to equal-weights (i.e. maximum $w_r \rightarrow 1/13 $) is more advantageous during the summer monsoon over the Tibetan Plateau and during the winter over Hindu Kush.}
\label{fig:max_weights}
\end{figure}

\clearpage

\section{MoE and BCM CRPS differences}

\begin{figure}[ht]
\centering
\includegraphics[width=\textwidth]{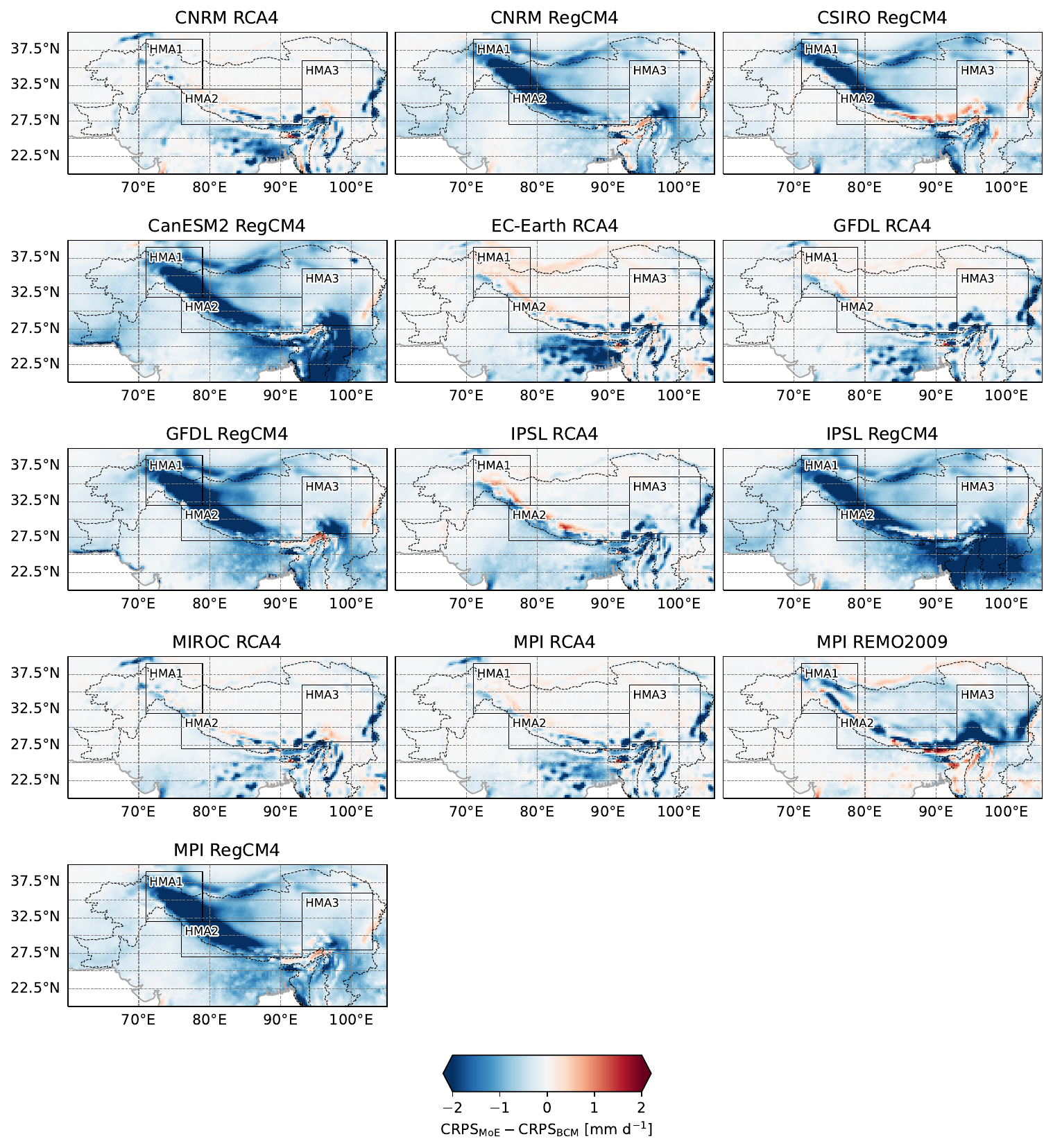}
\caption[Annual MoE and BCM CRPS differences for validation period]{Annual CRPS difference between the MoE and RCM surrogates (BCM) for the held-out validation period (1981--2005). The MoE outputs are generated from $10^5$ RCM surrogate samples for each month and location. Negative values (blue) imply MoE matches APHRODITE more closely while positive values (red) represent times and locations where the BCM performs better. Overall, the MoE outperforms the BCMs over the entire year with large improvement over HMA1 and HMA2, especially for the RegCM4 surrogates.}
\label{fig:moe_vs_rcms}
\end{figure}
\clearpage

\begin{figure}[ht]
\center
\includegraphics[width=\textwidth]{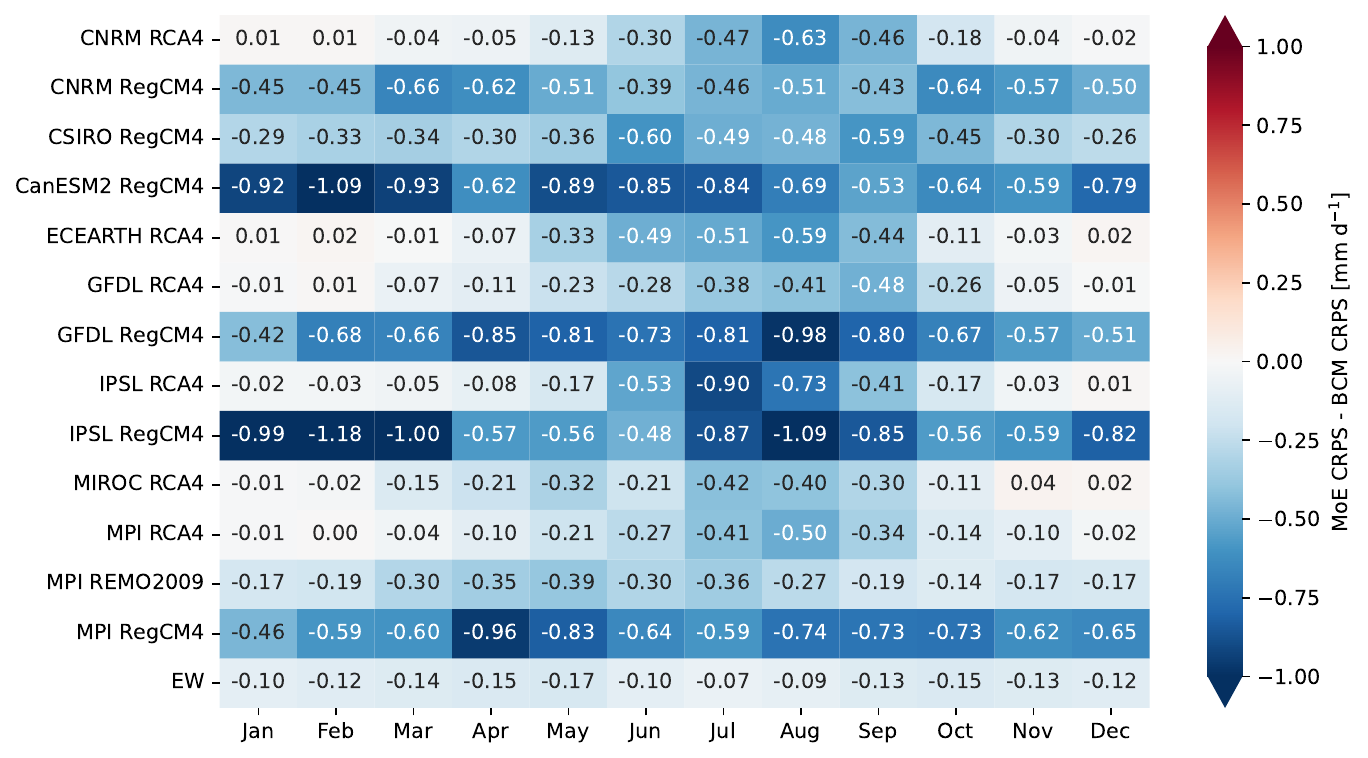}
\caption[CRPS scorecard for the MoE, BCM and EW models over validation period ]{Scorecard showing the differences between the CRPS yielded by the MoE and that of each RCM surrogate (BCM) as well as the EW over the held-out validation period (1981--2005).
Each cell represents the difference for a specific model and month, averaged over the entire spatial domain. The MoE and EW outputs are generated from $10^5$ RCM surrogate samples for each month and location.
Positive values (red) indicate higher CRPS for the MoE compared to the BCM, while negative values (blue) indicate lower CRPS for the MoE compared to the BCM.}
\label{fig:scorecard}
\end{figure}

\clearpage

\section{Historical predictions biases}
\label{app:hist_pred_abs}

\begin{figure}[ht]
\center
\includegraphics[width=\textwidth]{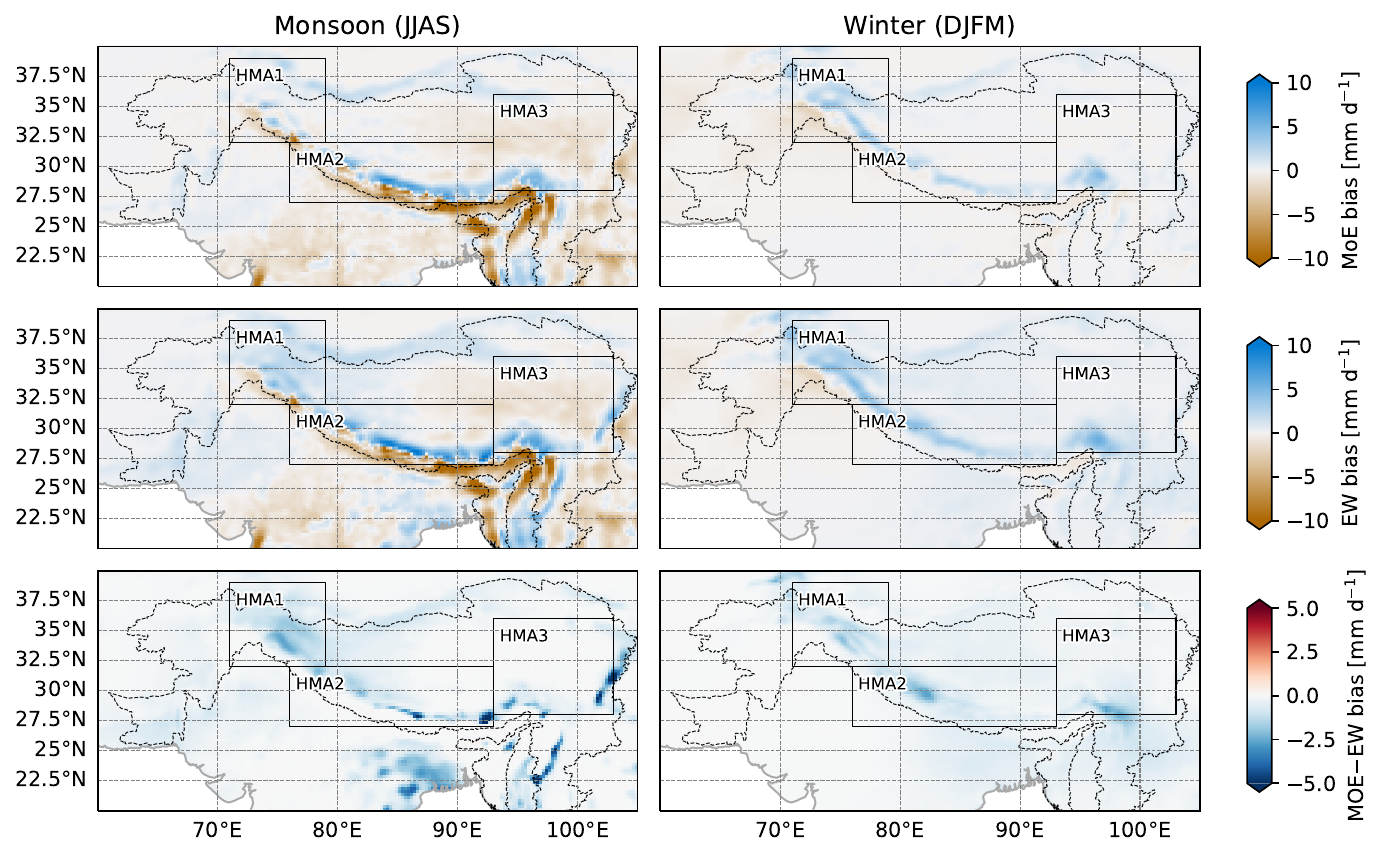}
\caption[Historical MoE and EW bias with respect to APHRODITE]{Historical MoE and EW bias for median precipitation with respect to APHRODITE over HMA. The MoE bias (top), the EW bias (middle), and the bias difference between MoE and EW (bottom) are plotted for the historical reference period (1976--2005) for the summer monsoon (left) and winter (right). The MoE and EW outputs are generated from $10^5$ RCM surrogate samples for each month and location. The MoE makes large improvements over the EW, in particular, over locations where precipitation is overestimated by the EW.}
\label{fig:moe_hist_abs}
\end{figure}

\clearpage

\section{Near-future predictions}
\label{app:near_future}

\begin{figure}[ht]
\center
\includegraphics[width=\textwidth]{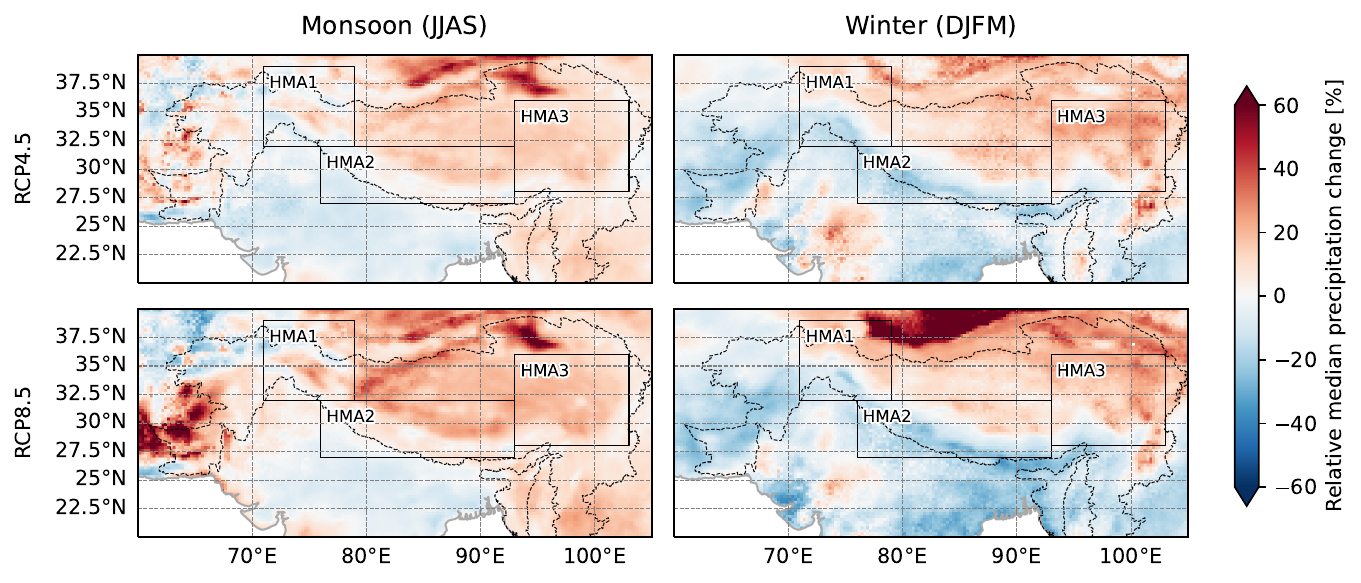}
\caption[Relative changes between near-future and historical MoE precipitation medians for RCP4.5 and RCP8.5]{Relative changes between near-future (2036--2065) and historical (1976--2005) MoE precipitation across HMA. The plot shows the median changes for RCP4.5 (top) and RCP8.5 (bottom) during the summer monsoon (left) and winter (right). The MoE and EW outputs are generated from $10^5$ RCM surrogate samples for each month and location. Results are similar to far-future predictions with less pronounced changes.}
\label{fig:nf_moe_preds}
\end{figure}

\begin{figure}[ht]
\center
\includegraphics[width=\textwidth]{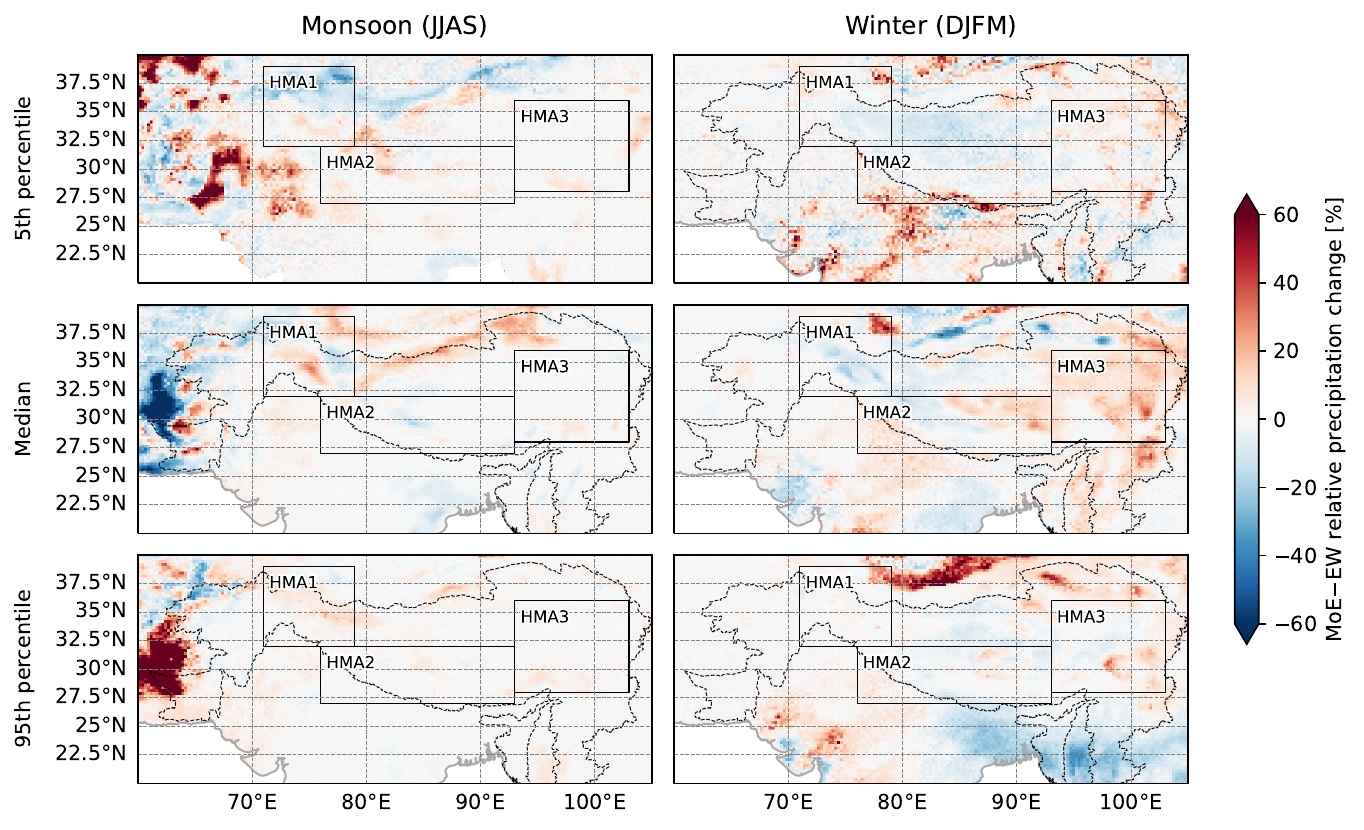}
\caption[MoE and EW relative prediction differences for near-future under RCP8.5]{MoE and EW relative prediction differences for the near future (2066--2095) under RCP8.5 across HMA with respect to their historical reference predictions. The plot shows the difference between the predicted relative changes for the 5\textsuperscript{th} percentile (bottom), median (middle), and 95\textsuperscript{th} percentile (top) during the summer monsoon (left) and winter (right). The MoE and EW outputs are generated from $10^5$ RCM surrogate samples for each month and location. Results are similar to far-future predictions with less pronounced differences.}
\label{fig:nf_pred_diffs}
\end{figure}

%TC:endignore
\end{document}